\documentclass[aps,prl,10pt,floatfix,superscriptaddress,noeprint,reprint]{revtex4-2}
\usepackage{amsmath,amsfonts,amssymb,physics,graphicx}
\newcounter{mynote}
\usepackage{comment}
\usepackage[dvipsnames]{xcolor}
\usepackage[colorlinks=true,
hyperfootnotes=true,
bookmarks=true,
breaklinks=true,
citecolor=blue,
urlcolor=blue,
linkcolor=blue,
]{hyperref}
\usepackage[normalem]{ulem}
\usepackage{soul}
\sethlcolor{yellow}

\newcommand{\como}{Center for Nonlinear and Complex Systems, Dipartimento di Scienza e Alta Tecnologia, Università degli Studi dell'Insubria, Via Valleggio 11, 22100 Como, Italy}
\newcommand{\infnM}{INFN, Sezione di Milano, 20133 Milano, Italy}
\newcommand{\genova}{Dipartimento di Fisica, Università di Genova, Via Dodecaneso 33, I-16146 Genova, Italy}
\newcommand{\CNR}{CNR-SPIN, Via Dodecaneso 33, I-16146 Genova, Italy}

\begin{document}
\title{Initial-Slip Dynamics Enables the Quantum Mpemba Effect}
\author{Maristella Crotti}
\affiliation{\como}
\affiliation{\infnM}
\author{Fabio Cavaliere}
\affiliation{\genova}
\affiliation{\CNR}
\author{Dario Ferraro}
\affiliation{\genova}
\affiliation{\CNR}
\author{Giuliano Benenti}
\affiliation{\como}
\affiliation{\infnM}
\author{Maura Sassetti}
\affiliation{\genova}
\affiliation{\CNR}

\date{\today}

\begin{abstract} 
We investigate the quantum Mpemba effect, whereby a system initially farther from equilibrium relaxes faster than one initially closer to equilibrium, emerging in a quantum harmonic oscillator linearly coupled to a bosonic environment.
Starting from factorized thermal states at different temperatures, we derive exact analytical criteria for the occurrence of the quantum Mpemba effect valid also beyond the typically considered weak-coupling and Markovian approximations. Most importantly, we show that the effect is enabled by the initial-slip dynamics generated by the sudden switch-on of the system--bath interaction. Neglecting this contribution completely suppresses the anomalous relaxation, identifying transient system--bath correlations as the key ingredient underlying the quantum Mpemba effect.
\end{abstract}

\maketitle

\textit{Introduction.}--- 
The relaxation of open quantum systems toward thermal equilibrium is a fundamental problem in nonequilibrium statistical mechanics \cite{Petruccione}. Understanding how a quantum system exchanges energy and information with its environment is central to quantum thermodynamics~\cite{vinjanampathy2016quantum, Binder2018,Campbell_2026}, quantum information processing~\cite{BenentiCasati}, and the theory of dissipative quantum systems~\cite{Weiss2012, Rivas2014}. While relaxation is generally expected to take longer as the initial state moves farther from equilibrium, anomalous relaxation phenomena demonstrate that this intuition can fail.

Among such phenomena, a prominent example is given by the Mpemba effect~\cite{Mpemba1969,LuRaz,Lasanta2017,Bechhoefer2021}, namely the counterintuitive situation in which two systems initially prepared at different temperatures relax toward equilibrium at different rates, so that the hotter one thermalizes before the colder one. Although the classical Mpemba effect remains controversial~\cite{Burridge2016,Burridge2020} and continues to stimulate active research~\cite{Nava2025,Teza2026, Summer2026, Liu2026, Melles2026}, its quantum counterpart has recently attracted considerable attention~\cite{Ares2025}.
Previous studies have investigated the quantum Mpemba effect in Markovian~\cite{Manikandan2021,Carollo2021,Kochsiek2022,Amorim2023, Evander2023,Chatterjee2023,Moroder2024,Nava2024,Wang2024,Caldas2025, Biswas2025, Longhi2025,Medina2025,Zhang2025, Qian2025,Saliba2026,Avitan2026, Rinaldi26} and non-Markovian open-system dynamics~\cite{Strachan2025,Zhang2026,Wang2025,Chirico26}, in relaxation toward nonequilibrium steady states~\cite{Nava2024,Wang2024, Grazi2026, Zhao2026, Ge2026}, and in quenched closed-system dynamics~\cite{Ares2023,Joshi2024, Rylands2024}.

Despite these developments, a fundamental question remains open: can the quantum Mpemba effect emerge within exact open-system dynamics, beyond the conventional Born--Markov approximation, starting from simple thermal states, without engineering ad hoc coherences or specially prepared initial conditions?
More importantly, if so, what physical mechanism underlies its emergence?

In this Letter, we answer these questions by considering a quantum harmonic oscillator linearly coupled to a bosonic environment within the Caldeira--Leggett model~\cite{CaldeiraLeggett,Caldeira1983, Grabert1988,Ingold02, Weiss2012, Manara26}. This paradigmatic model of quantum dissipation admits an exact treatment through the quantum Langevin equation. Owing to the linearity of the dynamics, the reduced state of the oscillator remains Gaussian at all times and is therefore fully characterized by its covariance matrix~\cite{Serafini2017}. This framework enables us to determine the exact relaxation dynamics of the systems without invoking weak-coupling or Markovian approximations.

Specifically, we consider initially factorized thermal states, in which the oscillator is uncorrelated from the bath. By introducing a trace-distance measure based on the Frobenius norm for Gaussian states~\cite{Banchi2015, Link2015}, we derive exact criteria for the occurrence of the direct quantum Mpemba effect between two initial thermal states of the system at different temperatures, $T_{\rm c}<T_{\rm h}$. We identify two distinct dynamical regimes: a weak-coupling underdamped (UD) regime, where the effect is reentrant and oscillatory, and overdamped (OD) moderate-coupling and UD strong-coupling regimes~\cite{Einsiedler2020}, where it becomes robust and is governed by a sharp condition on the amplitude of the slowest relaxation mode. More importantly, we show that the quantum Mpemba effect is enabled by the transient system--bath correlations generated immediately after the sudden switch-on of the interaction and encoded in the initial-slip term of the quantum Langevin equation~\cite{Ford1988, Hu1992}. Neglecting this contribution completely suppresses the effect, revealing the long time persistence of the initial-slip dynamics as the key mechanism underlying the quantum Mpemba effect.
More broadly, our results highlight the importance of transient system--bath correlations in quenched open-system dynamics and motivate the investigation of more general, initially correlated system--bath states in 
quantum relaxation processes.

\textit{Model and covariance matrix evolution.}---
We consider a quantum harmonic oscillator coupled to a bosonic reservoir~\cite{CaldeiraLeggett,Ingold02,Weiss2012}, described by the total Hamiltonian $H = H_{\rm osc} + H_{\rm res} + H_{\rm int}$, where 
$H_{\rm osc} = \frac{p^2}{2m} + \frac{1}{2}m\omega_0^2 x^2$, 
$H_{\rm res} = \sum_k \left( \frac{p_k^2}{2m_k} + \frac{1}{2}m_k\omega_k^2 x_k^2 \right)$, and
$H_{\rm int} = -x \sum_k c_k x_k + \frac{1}{2}x^2 \sum_k \frac{c_k^2}{m_k\omega_k^2}$. 
Here, $m$ and $\omega_0$ are the mass and the frequency of the system oscillator, $m_k$ and $\omega_k$ are the masses and the frequencies of the modes composing the bosonic bath, while $c_k$ represents the coupling strength between the system and the $k$-th reservoir mode. The last term in $H_{\rm int}$
is the usual potential-renormalization counter-term~\cite{CaldeiraLeggett,Ingold02,Weiss2012, Vool17, Cavaliere2025, Cavaliere2026, Capone26}.

We assume an initially factorized state $\rho(0) = \rho_{\rm res} \otimes \rho_{\rm osc}$, where $\rho_{\rm res} = e^{-H_{\rm res}/k_B T} / Z_{\rm res}$ and $\rho_{\rm osc} = e^{-H_{\rm osc}/k_B T_\alpha} / Z_{\rm osc}$ are thermal states of the reservoir and the oscillator, respectively. Here, $k_B$ is the Boltzmann constant, while $Z_{\rm res}$ and $Z_{\rm osc}$ denote the respective partition functions. The bath temperature $T$ and the initial oscillator temperature $T_\alpha$ are completely independent of each other.

By defining the dimensionless quadratures $Q = \sqrt{m\omega_0}x$ and $P = p/\sqrt{m\omega_0}$ (setting $\hbar=1$ henceforth), the system dynamics can be derived from the Heisenberg equations of motion for the system and reservoir operators. This yields the quantum Langevin equation
\begin{equation}
\ddot{Q}(t) + \omega_0^2 Q(t) + \int_0^t d\tau\,\gamma(t-\tau)\dot{Q}(\tau) + \gamma(t)Q(0) = {\sqrt{\frac{\omega_0}{m}}}\xi(t),
\label{eq:Langevin}
\end{equation}
where $\gamma(t)=\gamma_0\omega_{\mathrm{D}}e^{-\omega_{\mathrm{D}}t}\vartheta(t)$ is the damping kernel, with $\gamma_0$ the oscillator–reservoir coupling, $\omega_D$ the cut--off frequency, $\vartheta(t)$ is the Heaviside step function and $\xi(t)$ represents the fluctuating noise operator originating from the bath initial conditions, satisfying $\langle \xi(t)\rangle = 0$ [see the Supplemental Material (SM) for details].

Due to the linearity of the dynamics, the state of the oscillator is fully characterized by its covariance matrix (CM)\cite{Cavaliere2025}\footnote{Here, the quantum expectation value $\langle O(t)\rangle$ is defined as $\tr\{\rho(0)O(t)\}$, where $O(t)$ is the operator evolved in time with respect to the total Hamiltonian $H$, and $\rho(0)$ is the total density matrix at $t=0$.}
\begin{equation}
\sigma_\alpha(t)
=\begin{pmatrix}
\langle Q^2(t)\rangle & \langle \{Q(t), P(t)\}\rangle/2 \\[1ex]
\langle \{Q(t), P(t)\}\rangle/2  & \langle P^2(t)\rangle
\end{pmatrix}.
\label{sigma}
\end{equation}
The formal solution of Eq.~\eqref{eq:Langevin} for the CM can be decomposed into a homogeneous part and a thermal fluctuation part~\cite{Cavaliere2025},
$\sigma_\alpha(t) = \sigma_{\rm h,\alpha}(t) + \sigma_{\rm th}(t)$.
The homogeneous contribution is given by 
\begin{equation}
    \sigma_{\rm h, \alpha}(t) = X(t) \sigma_\alpha(0) X^T(t),\ \ X(t) = \begin{pmatrix} \dot{\chi}(t) & \omega_0 \chi(t) \\
\ddot{\chi}(t)/\omega_0 & \dot{\chi}(t) \end{pmatrix},
\label{eq:Propagator}
\end{equation}
where $\sigma_\alpha(0) = \frac{1}{2} C_\alpha \mathbb{I}$ represents the initial CM, with $C_\alpha \equiv \coth[\omega_0/(2k_B T_\alpha)]$ and $\mathbb{I}$ the $2\times 2$ identity matrix.

The response function $\chi(t)$ is determined by the 
reservoir spectral density under consideration. From now on, we focus on the Drude spectral density $J(\omega) = m\gamma_0\omega_D^2 \frac{\omega}{\omega^2+\omega_D^2}$~\cite{Ingold02,Weiss2012}. The extension of our results to a broad class of spectral densities is presented in the SM. 
For the Drude spectral density, the response function can be expressed as $\chi(t) = \sum_{j=1}^3 \chi_j e^{-\lambda_j t}$\cite{Cavaliere2025}\footnote{This expression assumes the generic case in which the roots of the characteristic polynomial, $\mathcal{P}(\lambda) = 0$, are distinct (i.e., all poles are simple).}, 
where $\{\lambda_j\}$ are the roots (with positive real part $\Re[\lambda_j]>0$) of the characteristic polynomial 
$\mathcal{P}(\lambda) = \lambda^3 - \omega_D\lambda^2+(\omega_0^2 + \gamma_0\omega_D)\lambda-\omega_D\omega_0^2 = 0$
and the coefficients 
$\chi_j = (\omega_D - \lambda_j) \prod_{\ell \neq j} (\lambda_\ell - \lambda_j)^{-1}$.

The thermal contribution $\sigma_{\rm th}(t)$  is given by
\begin{equation}
\begin{split}
\langle Q^2(t)\rangle_{\mathrm{th}}
&=
\frac{\omega_0}{m}
\int_0^t d\tau_2
\int_0^t d\tau_1
\chi(\tau_1)\chi(\tau_2)
L(\tau_2-\tau_1),\\
\langle P^2(t)\rangle_{\mathrm{th}}
&=
\frac{1}{m\omega_0}
\int_0^t d\tau_2
\int_0^t d\tau_1
\dot\chi(\tau_1)\dot\chi(\tau_2)
L(\tau_2-\tau_1)\,.
\end{split}
\label{eq:thermal_general}
\end{equation}
Here, $L(t)$ is the symmetric part of the bath correlation function $\langle\xi(\tau_1)\xi(\tau_2)\rangle $~\cite{Cavaliere2025}, which, for the Drude spectral density, is mapped onto an infinite sum of exponentials, $L(t) = \sum_{n=0}^\infty L_n e^{-A_n |t|}$, where $A_0 = \omega_D$ and $A_{n\ge 1} = 2\pi nk_BT$ are the Matsubara frequencies~\cite{Weiss2012, FetterWalecka1971}. The specific form of the amplitudes $\{L_n\}$ and the detailed derivation are provided in the SM.

Substituting the above expression for $L(t)$ into Eq.~\eqref{eq:thermal_general} yields the exact expressions for the thermal contributions to the variances,
\begin{align}
\langle Q^2(t) \rangle_{\rm th} &= \sum_{i,j=1}^3 \sum_{n=0}^\infty \chi_i \chi_j L_n I_{ijn}(t), \label{eq:Q_exact} \\
\langle P^2(t) \rangle_{\rm th} &= \frac{1}{\omega_0^2}\sum_{i,j=1}^3 \sum_{n=0}^\infty (\lambda_i \chi_i)(\lambda_j \chi_j) L_n I_{ijn}(t), \label{eq:P_exact}
\end{align}
where
$I_{ijn}(t)=I_{ijn}(\infty)+\Delta I_{ijn}(t)$
is naturally decomposed into stationary and transient contributions,
\begin{align}
I_{ijn}(\infty)
&=
\frac{2\omega_0}{m(\lambda_i+\lambda_j)(\lambda_i+A_n)},
\label{eq:I_inf}
\\
\Delta I_{ijn}(t)
&=
\Delta I_{ijn}^{(1)}(t)+\Delta I_{ijn}^{(2)}(t)\nonumber\\
&=
\frac{2\omega_0/m}{A_n-\lambda_j}
\left[
\frac{e^{-(\lambda_i+A_n)t}}{\lambda_i+A_n}
-
\frac{e^{-(\lambda_i+\lambda_j)t}}{\lambda_i+\lambda_j}
\right].
\label{eq:I_trans}
\end{align}
Consequently, the thermal contributions to the variances inherit the same decomposition into a stationary component and two transient contributions ($S\in\{Q,P\}$)
\footnote{The off-diagonal element of the CM follows from $\frac{1}{2}\langle \{Q(t), P(t)\}\rangle = \frac{1}{2\omega_0}\frac{d}{dt}\langle Q^2(t)\rangle$.}\setcounter{mynote}{\value{footnote}}:
\begin{equation}
\!\!\!\langle S^2(t) \rangle_{\rm th}
=
\langle S^2(\infty) \rangle_{\rm th}
+
\langle \Delta S^{(1)}(t)\rangle_{\rm th}
+
\langle \Delta S^{(2)}(t)\rangle_{\rm th}.
\label{eq:thermal_split}
\end{equation}
The details of the derivation are provided in the SM.

\textit{Relaxation to equilibrium and the Mpemba effect.}---
The exact solution derived above provides the complete time evolution of the reduced state. To characterize the relaxation toward the stationary state, we consider the Frobenius distance~\cite{Serafini2017}
$
\mathcal{D}_\alpha(t)
=
\sqrt{\tr\left\{\left[\rho_\alpha(t)-\rho(\infty)\right]^2\right\}},
$
between the reduced density matrix $\rho_\alpha(t)$ of the oscillator at time $t$ and its asymptotic state $\rho(\infty)$, whose CM is denoted by $\sigma(\infty)$. In the long-time limit,
\begin{equation}
\mathcal{D}_{\alpha}(t)
\approx
\frac{1}{4 D_{\infty}^{1/4}}
\sqrt{\tr^2\{y_{\alpha}(t)\}
+2\,\tr\{y_{\alpha}^2(t)\}},
\label{Dtraccia}
\end{equation}
where $D_{\infty}=\det\sigma(\infty)$ and
$y_{\alpha}(t)=\sigma(\infty)^{-1}\Delta\sigma_{\alpha}(t)$,
with
$\Delta\sigma_\alpha(t)\equiv\sigma_\alpha(t)-\sigma(\infty)$
(see SM).
We emphasize that similar results are obtained when the Gaussian infidelity is used as an alternative measure.

We use the Frobenius distance to quantify the  Mpemba effect. Consider two initial thermal states of the oscillator prepared at temperatures $T_{\mathrm c}$ (``cold'') and $T_{\mathrm h}$ (``hot''). A direct Mpemba effect occurs, for
$T<T_{\mathrm c}<T_{\mathrm h}$,
when the initially hotter state relaxes faster than the initially colder one. 
At finite system--bath coupling, however, the temperature ordering alone does not guarantee the required
initial ordering,
$\mathcal{D}_{\rm h}(0)>\mathcal{D}_{\rm c}(0)$.
A sufficient condition ensuring this ordering is discussed in the SM.
In all the examples presented below, the initial temperatures are chosen
accordingly, so that 
$\mathcal{D}_{\rm h}(0)>\mathcal{D}_{\rm c}(0)$.
We define the estimator
$
\Delta\mathcal{D}(t)
=
\mathcal{D}_{\mathrm c}(t)-\mathcal{D}_{\mathrm h}(t).
$
Given the initial ordering,
$\Delta\mathcal{D}(0)<0$
the direct Mpemba effect is characterized by the asymptotic
inversion of the distance ordering, i.e.,
$\Delta\mathcal{D}(t)>0$ at long times.
This means that the initially hotter state becomes asymptotically
closer to the stationary state than the initially colder one.

The long-time behavior of $\Delta\mathcal{D}(t)$ is determined by the asymptotic relaxation dynamics. Assuming the ordering
$\Re(\lambda_3)\leq\Re(\lambda_2)\leq\Re(\lambda_1)$,
the long-time evolution is governed by the characteristic root, or complex-conjugate pair of roots, entering $\chi(t)$ with the smallest positive real part. This corresponds to the slowest relaxation mode and determines the asymptotic exponential approach to the stationary state. 
From now on, we restrict our analysis to the regime
$
k_B T>\mathrm{Re}(\lambda_3)/2\pi,
$
where the contribution $\Delta I_{ijn}^{(2)}(t)$ governs the long-time dynamics.
Indeed, Viète's relation for the characteristic cubic polynomial
$\mathcal{P}(\lambda)$,
$
\sum_{i=1}^{3}\lambda_i=\omega_D,
$
implies
$
\mathrm{Re}(\lambda_3)<\omega_D=A_0.
$
Moreover, the Matsubara frequencies satisfy
$
A_{n\ge1}=2\pi n k_B T,
$
so that, in the temperature regime under consideration,
$
A_n\ge A_1=2\pi k_B T>\mathrm{Re}(\lambda_3).
$
Therefore, the exponential contributions to
$\Delta I_{ijn}^{(1)}(t)$ decay faster than those of
$\Delta I_{ijn}^{(2)}(t)$ and are subleading in the long-time limit. The regime $k_B T<\mathrm{Re}(\lambda_3)/2\pi$ is discussed in the SM.

\textit{UD weak-coupling regime.}---
We first analyze the UD weak-coupling regime ($\gamma_0<2\omega_0$ and $\gamma_0,\omega_0\ll\omega_D$), where the slowest characteristic roots form a complex-conjugate pair \cite{Cavaliere2025},
$\lambda_{2,3}=\lambda\mp i\eta$,
with corresponding weights satisfying $\chi_2=\chi_3^*$. The remaining root is real and satisfies $\lambda_1\gg\lambda$, yielding only exponentially suppressed contributions at long times.
The asymptotic behavior of the response function therefore takes the form
$
\chi(t)=
e^{-\lambda t}
\left[
\chi_3 e^{-i\eta t}
+\chi_3^* e^{i\eta t}
\right].
$
The homogeneous contribution to the CM then becomes
$\sigma_{\rm h, \alpha}(t)=e^{-2\lambda t}\tilde{\sigma}_{\rm h,\alpha}(t)$,
where $\tilde{\sigma}_{\rm h,\alpha}(t)$ follows directly from Eq.~\eqref{eq:Propagator}.
Retaining only the dominant term $\Delta I_{ijn}^{(2)}(t)$ and the slowest roots $\lambda_{2,3} = \lambda \mp i\eta$, we obtain $\langle \Delta Q^{(2)}(t)\rangle_{\mathrm{th}} =B_{\rm th, 1}(t)e^{-2\lambda t}$ and $\langle \Delta P^{(2)}(t)\rangle_{\mathrm{th}} =B_{\rm th, 2}(t)e^{-2\lambda t}$, where $B_{\rm th, 1}(t)$ and $B_{\rm th, 2}(t)$ are oscillatory functions of time and are provided in the End Matter  \footnotemark[\value{mynote}].
Consequently, the asymptotic CM for an initial oscillator temperature $T_\alpha$ can be written as $\sigma_\alpha(t)=\sigma(\infty)+\Delta\sigma_\alpha(t)$, with
$
\Delta\sigma_\alpha(t)
=
\sigma_{{\rm h},\alpha}(t)+\Delta\sigma^{(2)}_{\rm th}(t)
=
e^{-2\lambda t}\mathbb{D}_\alpha(t).
$ Here, $\sigma(\infty)$ is the stationary CM associated with $I_{ijn}(\infty)$, while $\mathbb{D}_\alpha(t)$ collects the prefactors multiplying the common exponential decay $e^{-2\lambda t}$; for example,
$
\mathbb{D}^{(1,1)}_\alpha(t)
=
\tilde{\sigma}^{(1,1)}_{{\rm h},\alpha}(t)+B_{{\rm th},1}(t),
$
while the other matrix elements are defined analogously.

\begin{figure}[ht]
\centering
\includegraphics[height=0.18\textwidth]{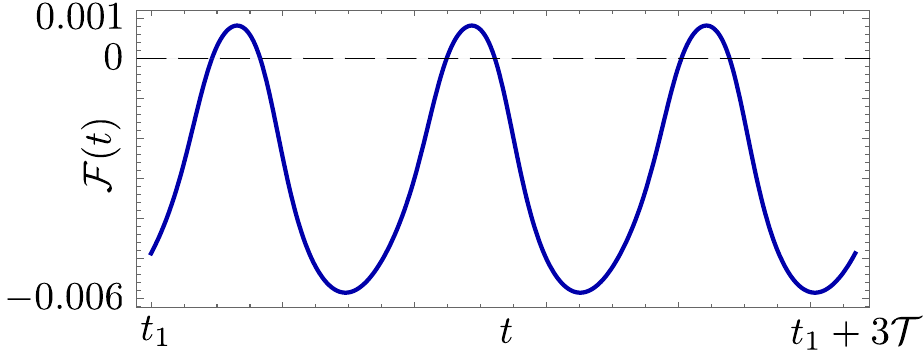}
\caption{Plot of ${\cal F}(t)$ as a function of $t$ (unit $\omega_0^{-1}$) for $\gamma_0=\omega_0$, $\omega_{\mathrm{D}}=20\omega_0$, $k_{\mathrm{B}}T=10\omega_0/(2\pi)$, $k_{\mathrm{B}}T_{\mathrm{c}}=1.85\omega_0$ and $k_{\mathrm{B}}\Delta T\equiv k_B(T_{\rm h}-T_{\rm c})=0.01\omega_0$. Here, $\mathcal{T}=\pi/\eta$. }
\label{fig:FigDWeak}
\end{figure}

It follows that 
$y_\alpha(t)=e^{-2\lambda t}\sigma^{-1}(\infty)\mathbb{D}_\alpha(t)\equiv e^{-2\lambda t}\mathbb{E}_\alpha(t)$, and, using Eq. \eqref{Dtraccia},
\begin{align}
\Delta\mathcal{D}(t)&=\frac{e^{-2\lambda t}}{4 D_{\infty}^{1/4}}\big[
\sqrt{\tr^2\{\mathbb{E}_{\rm c}(t)\}
+2\,\tr\{\mathbb{E}^2_{\rm c}(t)\}}\nonumber\\
&-\sqrt{\tr^2\{\mathbb{E}_{\rm h}(t)\}
+2\,\tr\{\mathbb{E}^2_{\rm h}(t)\}}\big]\equiv e^{-2\lambda t}\mathcal{F}(t).
\end{align}
Here $\mathcal{F}(t)$ is a periodic function, with period $\mathcal{T}=\pi/\eta$. 
Figure~\ref{fig:FigDWeak} shows the long-time behavior of $\mathcal{F}(t)$ for representative parameters.
The periodic sign changes of $\mathcal{F}(t)$ lead to alternating time windows in which $\mathcal{F}(t)>0$ and the direct Mpemba effect occurs.
As a result, the effect is reentrant, repeatedly emerging and disappearing as the dynamics evolve.
Furthermore, due to the weak coupling, the overall amplitude of the effect is strongly suppressed and vanishes in the uncoupled limit $\gamma_0\to0$.  

\textit{OD moderate-coupling and UD strong-coupling regimes.}--- 
In the former regime, defined by $2\omega_0 < \gamma_0 < \omega_D/4$, all three characteristic roots are real, with $\lambda_3$ being the slowests. In the latter, where $\omega_0 < \omega_D < 4\gamma_0$, the slowest root $\lambda_3$ remains real, whereas the other two roots form a complex-conjugate pair with significantly larger real parts~\cite{Cavaliere2025}. 

In the long-time limit, for both regimes the response function therefore reduces to $\chi(t) \simeq \chi_3 e^{-\lambda_3 t}$ (with $\chi_3 > 0$ \footnote{
The weight $\chi_3$ is real and positive in both regimes. In the OD regime this follows from the
fact that all characteristic roots are real and ordered as
$\lambda_3<\lambda_{1,2}$, with $\lambda_3<\omega_D$. In the
UD strong-coupling regime, $\lambda_1$ and $\lambda_2$ form a
complex-conjugate pair, so that
$
\lambda_1+\lambda_2=2\mathrm{Re}(\lambda_2),
\,
\lambda_1\lambda_2=|\lambda_2|^2 ,
$
and the denominator of $\chi_3$ is strictly positive (its discriminant,
$
\Delta=
-4\mathrm{Im}(\lambda_2)^2<0 ,
$
is negative, while the leading coefficient is positive). Since
$\lambda_3<\omega_D$, the numerator is also positive, proving
$\chi_3>0$.
}). 
Accordingly, the homogeneous contribution to the CM becomes
$
\sigma_{\rm h,\alpha}(t) = B_{\alpha} e^{-2\lambda_3 t} \mathbb{P},
$
where $B_{\alpha} = \frac{1}{2} C_\alpha \chi_3^2 \left(\omega_0^2 + \lambda_3^2\right)$ and
\begin{equation}
\mathbb{P} = \begin{pmatrix} 
1 & -\lambda_3/\omega_0 \\ 
-\lambda_3/\omega_0 & (\lambda_3/\omega_0)^2 
\end{pmatrix}.
\end{equation}
The dominant thermal contribution in the $k_BT>\lambda_3/2\pi$ regime is obtained by retaining the leading term in the asymptotic expansion, namely $i=j=3$ in $\Delta I_{ijn}^{(2)}(t)$.
Accordingly, 
$\langle \Delta Q(t)\rangle_{\rm th}^{(2)}\simeq B_{\rm th} e^{-2\lambda_3 t}$, where the coefficient $B_{\rm th}$ is reported in the End Matter. 
Similarly,  
$\langle \Delta P(t)\rangle_{\rm th}^{(2)} \simeq \left({\lambda_3}/{\omega_0}\right)^2 B_{\rm th} e^{-2\lambda_3 t}$.

Collecting the above results, the asymptotic CM can be written as $\sigma_\alpha(t)
=
\sigma(\infty)
+
\Delta\sigma_\alpha(t)$, where
$
\Delta\sigma_\alpha(t)
=
\sigma_{{\rm h},\alpha}(t)
+
\Delta\sigma_{\rm th}^{(2)}(t),
$
and $\sigma(\infty)$ is 
explicitly evaluated in Ref.~\cite{Cavaliere2025, Weiss2012}.
The time-dependent part of the CM takes the compact form $\Delta\sigma_{\alpha}(t) = e^{-2\lambda_3 t} \left[B_\alpha + B_{\mathrm{th}}\right] \mathbb{P}$.

Therefore, $y_{\alpha}(t)=\sigma^{-1}(\infty)\Delta\sigma_{\alpha}(t) \approx e^{-2\lambda_3 t} \left[B_\alpha + B_{\mathrm{th}}\right] \mathbb{M}_0$, where $\mathbb{M}_0=\sigma(\infty)^{-1}\mathbb{P}$.
Substituting this expression into Eq.~\eqref{Dtraccia}, we obtain
\begin{equation}
\Delta \mathcal{D}(t) = \frac{\sqrt{3}}{4D_{\infty}^{1/4}} \mathcal{C}_0 \Big[ |B_{\mathrm{c}}+B_{\mathrm{th}}| - |B_{\mathrm{h}}+B_{\mathrm{th}}| \Big] e^{-2\lambda_3 t},
\label{eq:DeltaDHighT}
\end{equation}
where $\mathcal{C}_0=\left|\tr(\mathbb{M}_0)\right| > 0$.

Therefore, the asymptotic inversion occurs when 
$\Delta B=|B_{\mathrm{c}}+B_{\mathrm{th}}| - |B_{\mathrm{h}}+B_{\mathrm{th}}|>0$, which is equivalent to (see End Matter) \begin{equation}
B_{\mathrm{th}} < -\frac{1}{2}(B_{\mathrm{c}}+B_{\mathrm{h}})\,.
\label{eq:MpembaCondition}
\end{equation}
For a given pair of initial temperatures $T_{\mathrm c}$ and
$T_{\mathrm h}$, Eq.~\eqref{eq:MpembaCondition} provides the necessary
and sufficient condition for the asymptotic inversion $\mathcal D_{\rm h}(t)<\mathcal D_{\rm c}(t)$.

Note that, since the right-hand side of Eq. \eqref{eq:MpembaCondition} decreases monotonically with both $T_{\mathrm c}$ and $T_{\mathrm h}$ (see End Matter), 
the most restrictive condition for the existence of the asymptotic inversion is obtained in the limit $T_{\mathrm c}, T_{\mathrm h} \to T$, that is $B_{\mathrm{th}}(T) < -B_{\rm res}{(T)}$, with $B_{\rm res}(T)=\frac{1}{2} C_{\rm res}{(T)} \chi_3^2 \left(\omega_0^2 + \lambda_3^2\right)$ and $C_{\rm res}{(T)}= \coth[\omega_0/(2k_B T)]$. 

This condition is fulfilled in the OD moderate- and UD strong-coupling regimes, where $\lambda_3<\omega_0$, provided that
$T>T^*(\gamma_0,\omega_D)$, where the critical temperature $T^*$ is defined by the equality $B_{\mathrm{th}}({T^*}) = -B_{\rm res}{(T^*)}$ and marks the threshold above which the inversion becomes accessible. Remarkably,
$
T^*(\gamma_0,\omega_D)>
\frac{\lambda_3(\gamma_0,\omega_D)}{2\pi},
$
so that the onset of the effect always occurs in the temperature range considered.
It can be shown (see End Matter) that the critical temperature decreases monotonically with increasing system--bath coupling $\gamma_0$, thereby enlarging the parameter region where the inversion can occur, whereas increasing the cut-off frequency raises the critical
temperature. 

Moreover, whenever
$T>T^{*}{(\gamma_0,\omega_{\mathrm{D}})}$, there always exists a finite region in the
$(T_{\mathrm c},T_{\mathrm h})$ plane where  the asymptotic inversion occurs. To see this, let
$T_{\mathrm h}=T_{\mathrm c}+\Delta T$ with
$\Delta T>0$. When $T > T^{*}(\gamma_0,\omega_D)$, Eq.~\eqref{eq:MpembaCondition} is strictly satisfied in the limit $T_{\mathrm c} \rightarrow T$ and $\Delta T \rightarrow 0$. Since the right-hand side of
Eq.~\eqref{eq:MpembaCondition} decreases monotonically with both $T_{\mathrm c}$ and $\Delta T$, the inequality is not restricted to infinitesimal perturbations; rather, it remains satisfied over a finite, contiguous region of the $(T_{\mathrm c},\Delta T)$ plane before eventually breaking down when $T_{\mathrm c}$ or $\Delta T$ becomes too large.
\begin{figure}[ht]
\centering
\includegraphics[height=0.18\textwidth]{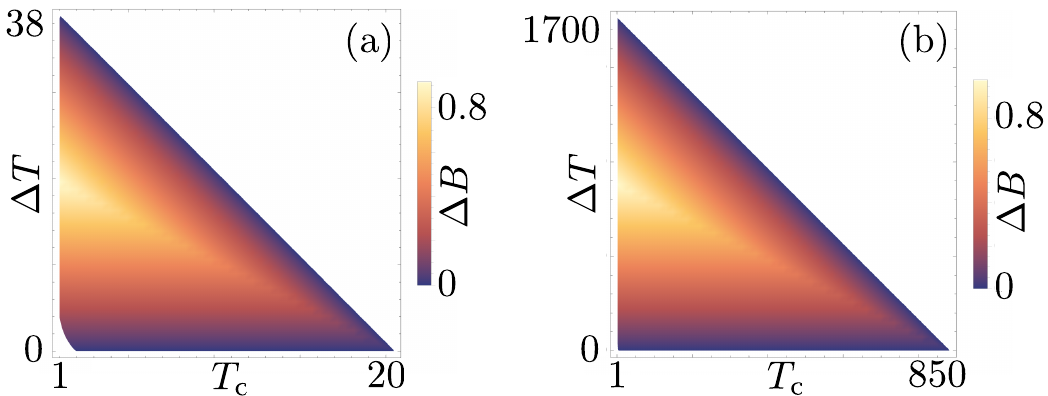}
\caption{Density plots of $\Delta B$ as a function of $T_{\mathrm c}$ and $\Delta T$ (units $k_{\mathrm B}^{-1}\omega_0$) for $\omega_{\mathrm D}=20\omega_0$, $k_{\mathrm B}T=\omega_0$, and (a) $\gamma_0=5\omega_0$ (OD moderate-coupling regime); (b) $\gamma_0=30\omega_0$ (UD strong-coupling regime). In both panels $T>T^{*}(\gamma_0,\omega_{\mathrm D})$. 
The colored regions correspond to $\Delta B>0$ and $\Delta\mathcal{D}(0)<0$, where the 
Mpemba effect occurs.
}
\label{fig:triangle}
\end{figure}
Notice that the additional requirement imposed by the
initial-distance ordering,
$\Delta\mathcal{D}(0)<0$,
may further restrict this region. However, as shown in the SM, this
requirement is compatible with the asymptotic-inversion condition.
Therefore, within the regimes considered here,
the direct Mpemba effect can always be realized by a suitable choice of $T$, $T_ {\rm c}$, and $T_{\rm h}$ satisfying both constraints.

To illustrate the structure of the Mpemba-effect region, we analyze
$\Delta B$, recalling that $\Delta B > 0$ is necessary for the occurrence of the Mpemba effect.
Figure~\ref{fig:triangle} shows density plots of $\Delta B$ in the $(T_{\mathrm c},\Delta T)$ plane for the two regimes under examination. The plots highlight the regions where $\Delta B>0$ and the initial-condition constraint 
$\Delta\mathcal{D}(0)<0$
is satisfied. 
The resulting region exhibits a characteristic triangular geometry driven by the competition between $T_{\mathrm c}$ and $\Delta T$: as the base temperature $T_{\mathrm c}$ increases, the maximum allowable temperature difference $\Delta T$ before the condition $\Delta B > 0$ is violated shrinks accordingly. Notably, the effect reaches its maximum intensity near the bisector of this triangular domain. 
Finally, a comparison between the panels reveals that increasing $\gamma_0$ systematically expands the accessible $(T_{\mathrm c},\Delta T)$ domain. This demonstrates that a stronger system--bath interaction enhances the direct Mpemba effect.

\begin{figure*}
    \centering
    \includegraphics[width=0.75\linewidth]{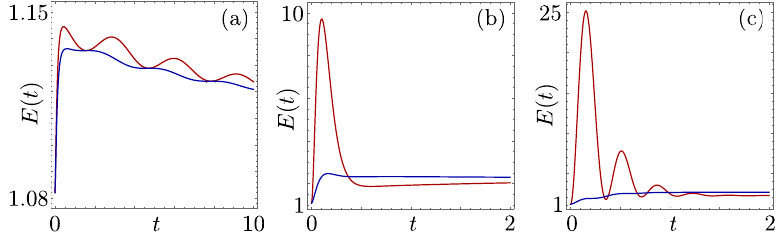}
    \caption{Time evolution of the oscillator energy,
$E(t)=\langle H_{\mathrm{osc}}(t)\rangle$ (in units of
$\omega_0$), with (red lines) and without (blue lines) the initial-slip contribution, for the three relaxation regimes considered in this work:
(a) UD weak-coupling, (b) OD moderate-coupling, and (c) and UD strong-coupling. Time is measured in units of $\omega_0^{-1}$.  The common parameters for all panels are
$k_{\mathrm{B}}T=k_{\mathrm{B}}T_{\mathrm{c},\mathrm{h}}=\omega_0$. The remaining
parameters are: (a) $\gamma_0=0.1\omega_0$,
$\omega_{\mathrm{D}}=20\omega_0$; (b) $\gamma_0=5\omega_0$,
$\omega_{\mathrm{D}}=20\omega_0$; and (c)
$\gamma_0=20\omega_0$, $\omega_{\mathrm{D}}=4\omega_0$.}
    \label{fig:energy}
\end{figure*}

\textit{Initial-slip and Mpemba effect.}---
Unlike previous studies on the quantum Mpemba effect~\cite{Longhi2025, Medina2025, Li2025, Zhang2025, Furtado2025, Riosmonje2026}, our analysis requires no specially engineered initial conditions. The oscillator is always prepared in a thermal state of the bare Hamiltonian, with no initial system--bath correlations, and the two initial conditions differ solely by their temperature, $T_{\rm h}$ and $T_{\rm c}$. 
This naturally raises the question of what physical mechanism gives rise to the anomalous relaxation observed here.

The origin of this behavior can be traced back to the non-adiabatic establishment of the system--bath interaction. Although the initial total state is factorized, the sudden switch-on of the coupling prevents the system from remaining in an instantaneous equilibrium configuration and generates system--bath correlations during the early stages of the evolution, whose effects survive also in the long time limit. In the exact quantum Langevin description (see Eq. \eqref{eq:Langevin}), these correlations are encoded in the initial slip contribution $\gamma(t)Q(0)$, which is absent in standard treatments based on an adiabatic switch-on or in commonly used Born--Markov approximations~\cite{Petruccione}.

This contribution is particularly relevant at short times, as illustrated in Fig.~\ref{fig:energy}, where we compare the oscillator energy in the three regimes discussed above with and without the initial-slip term. When the initial-slip is included, the oscillator energy exhibits pronounced transient oscillations and reaches values significantly larger than those attained in its absence. 
This enhancement becomes increasingly pronounced as the system--bath coupling strength increases and is particularly prominent in UD strong-coupling regime.
Thus, the initial-slip contribution induces a strong transient departure from equilibrium after the interaction is switched on
\footnote{We note that in a different context, the generation of excitations from the vacuum induced by a nonadiabatic modulation of the system parameters is known as the parametric dynamical Casimir effect~\cite{Dodonov_2010, Nation2012, dodonov2020physics}.}. The substantial energy enhancement displayed in Fig.~\ref{fig:energy} suggests that initial-slip dynamics may provide a promising route toward the fast charging of quantum batteries~\cite{Campaioli24,Ferraro2026natrevphys}.

Although the initial-slip contribution decays at long times, its effect is not limited to the short-time dynamics. Indeed, the correlations generated during this initial transient stage modify the subsequent relaxation pathway and leave a finite imprint on the asymptotic decay amplitudes. In the End Matter, we show that neglecting the initial-slip contribution suppresses the direct Mpemba effect, demonstrating that these correlations provide the relevant physical mechanism behind the anomalous relaxation observed here. Moreover, the Mpemba effect is also absent in the Born-Markov approximation \cite{Petruccione}, which neglects the initial-slip contribution (see SM), 
further corroborating the essential role of the initial correlations generated by the nonadiabatic switch-on of the system--bath interaction.

\textit{Conclusions.}---
In this Letter, we have demonstrated the existence of the quantum Mpemba effect
within the exact nonequilibrium dynamics of the Caldeira--Leggett model, deriving analytical conditions for its occurrence beyond weak-coupling and Markovian approximations. We identified distinct dynamical regimes in which the effect is either reentrant and oscillatory or robust and governed by an explicit condition on the amplitudes of the slowest relaxation mode.

Our central finding is that the quantum Mpemba effect is enabled by the transient dynamics generated by the sudden switch-on of the system--bath interaction. Here, correlations emerge due to the initial-slip term of the quantum Langevin equation, whose inclusion is essential for the emergence of the effect. Conversely, neglecting this contribution completely suppresses the anomalous relaxation.

More broadly, our results demonstrate that transient system--bath correlations generated during quenched open-system dynamics can have long-lasting consequences for relaxation toward equilibrium. They therefore motivate further investigations of nonequilibrium dynamics beyond the commonly assumed factorized system--reservoir state~\cite{Fleming2011} and suggest new approaches to controlling relaxation in open 
quantum systems.


\textit{Acknowledgments.}---
M.C. and G.B. acknowledge support from INFN through the project “QUANTUM”.

\bibliography{references}

\onecolumngrid

\newpage 
\section*{End Matter}    
\appendix
\twocolumngrid

\setcounter{equation}{0}
\renewcommand{\thesection}{A}
\textit{Appendix A: Details on the coefficients of the thermal contributions} ---
We provide here the explicit expressions of the thermal prefactors $B_{\rm th,1/2}$ for the UD weak coupling regime,
\begin{equation}
\begin{split}
B_{\rm th, 1}(t)=& \begin{aligned}[t]
- \frac{\omega_0}{m}  S_1(\lambda+i\eta)\Big[ \frac{|\chi_3|^2}{\lambda} + \frac{\chi_3^2\,e^{-2i\eta t}}{\lambda+i\eta}   \Big]+ \text{c.c.},
\end{aligned}  \\
B_{\rm th, 2}(t) =& \begin{aligned}[t]
- \frac{1}{m\omega_0\lambda}  S_1(\lambda+i\eta)\Big[ {|\chi_3|^2(\lambda^2+\eta^2)} 
\\+ \chi_3^2 \lambda(\lambda+i\eta) e^{-2i\eta t}  \Big]+ \text{c.c.},
\end{aligned}  
\end{split}
\label{eq:sig}
\end{equation}
and $B_{\rm th}$ for the OD moderate coupling and UD strong coupling regimes, $B_{\rm th} =-\frac{\omega_0\chi_3^2}{m\lambda_3}S_1(\lambda_3)$,
where $S_1(x)=\sum_{n=0}^{\infty}L_n/(A_n-x)=\frac{m \gamma_0 \omega_D}{\omega_D-\lambda_3}F(x)$, with
\begin{equation}
\begin{split}
F(x)=
\frac{\alpha}{2\pi}+\frac{\omega_Dx}{\pi(\omega_D+x)}\bigg[\psi\bigg(1-\frac{x}{\alpha}\bigg)-
\psi\bigg(1+\frac{\omega_D}{\alpha}\bigg)
\bigg],
\label{eq:BQ}
\end{split}
\end{equation}
where $\psi(x)$ is the digamma function and we defined $\alpha\equiv2\pi k_BT$. The explicit evaluation of $S_1(x)$ is reported in the SM.

The constraint involving $B_{\rm th}$, introduced in Eq.~\eqref{eq:MpembaCondition}, can be written explicitly as
\begin{equation}
B_{\mathrm{th}} < -\frac{\chi_3^2\left(\omega_0^2+\lambda_3^2\right)}{4} \left[ \coth\left(\frac{\omega_0}{2k_B T_{\mathrm{c}}}\right) + \coth\left(\frac{\omega_0}{2k_B T_{\mathrm{h}}}\right) \right].
\label{eq:MpembaCondition1}
\end{equation}
In the limit $T_{\rm c,h}\rightarrow T$, this reduces to
\begin{equation}
B_{\mathrm{th}}(T) < -\chi_3^2 \frac{\left(\omega_0^2+\lambda_3^2\right)}{2} \coth\left(\frac{\omega_0}{2k_B T}\right)\equiv f(T).
\label{eq:Mpemba2Condition}
\end{equation}
We now analyze this condition for the asymptotic inversion
in detail. 
For $k_{\mathrm{B}}T > \lambda_3/(2\pi)$, the arguments of the digamma function in Eq. \eqref{eq:BQ} are positive, making $B_{\mathrm{th}}(T)$ continuous and strictly decreasing (since $\psi'(x)>0$). Its asymptotic limits are
$
\lim_{T \to \lambda_3 / (2\pi k_{\mathrm{B}})} B_{\mathrm{th}}(T) = +\infty$ and  $\lim_{T \to \infty}B_{\mathrm{th}}(T) = -\mathcal{A}k_{\mathrm{B}}T,
$
where $\mathcal{A} = \chi_3^2 {\gamma_0\omega_0\omega_{\mathrm{D}}}/[{\lambda_3(\omega_{\mathrm{D}}-\lambda_3)}]$. Similarly, $f(T)$ is continuous, strictly decreasing, and asymptotically satisfies $f(T) \simeq -\mathcal{B}k_{\mathrm{B}}T$, with $\mathcal{B} = \chi_3^2 ({\omega_0^2+\lambda_3^2})/{\omega_0}$.
Using the properties of the digamma function \footnote{We use the strict monotonicity of $\psi(x)$ for $x>0$ and the identity $\psi(1+x)=\psi(x)+1/x$, yielding $\psi\left(1-{\lambda_3}/({2\pi k_{\mathrm{B}}T})\right) - \psi\left({\omega_{\mathrm{D}}}/({2\pi k_{\mathrm{B}}T})\right) < {2\pi k_{\mathrm{B}}T}/{\omega_{\mathrm{D}}}$.} and the identity $\coth x > 1/x$, one obtains $B_{\mathrm{th}}(T) > -\mathcal{A}k_{\mathrm{B}}T$ and $f(T) < -\mathcal{B}k_{\mathrm{B}}T$.
Consequently, if $\mathcal{A} \leq \mathcal{B}$, then
$
B_{\mathrm{th}}(T) > -\mathcal{A}k_{\mathrm{B}}T \geq -\mathcal{B}k_{\mathrm{B}}T > f(T),
$
which contradicts the condition (\ref{eq:Mpemba2Condition}), making the Mpemba effect impossible.
Since $\psi'(x) > 1/x^2$ and $\sinh x > x$, the derivatives satisfy $B'_{\mathrm{th}}(T) < -\mathcal{A}k_{\mathrm{B}}$ and $-\mathcal{B}k_{\mathrm{B}} < f'(T) < 0$. If $\mathcal{A} > \mathcal{B}$, then
$
B'_{\mathrm{th}}(T) < f'(T),
$
meaning that $B_{\mathrm{th}}(T) - f(T)$ is strictly decreasing. Because this difference varies from $+\infty$ (for $T \to \lambda_3/2\pi k_{\mathrm{B}}$) to $-\infty$ (for $T \to \infty$), there exists a unique temperature $T^*$ such that $B_{\mathrm{th}}(T^*) - f(T^*) = 0$. Thus, the asymptotic inversion condition is fulfilled for all $T > T^*$.

Using the characteristic equation identity $\gamma_0\omega_0\omega_{\mathrm{D}} = (\omega_{\mathrm{D}}-\lambda_3)(\omega_0^2+\lambda_3^2)$ \footnote{This identity follows from $\lambda_3$ being a root of the polynomial $\mathcal{P}(\lambda)$.}, the ratio of the coefficients simplifies to
$
{\mathcal{A}}/{\mathcal{B}} = {\omega_0}/{\lambda_3}.
$
The condition $\mathcal{A} > \mathcal{B}$ is therefore equivalent to $\lambda_3 < \omega_0$, which is always satisfied in the case of intermediate and strong coupling regimes with $\gamma_0>2\omega_0$.

\begin{figure}[ht]
\centering
\includegraphics[height=0.2\textwidth]{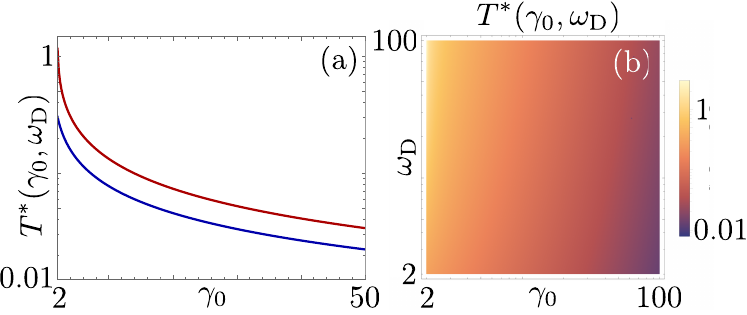}
\caption{(a) Critical temperature $T^{*}(\gamma_0,\omega_{\mathrm D})$ (units $k_{\mathrm B}^{-1}\omega_0$) as a function of $\gamma_0$ (units $\omega_0$) for $\omega_{\mathrm D}=20\omega_0$ (red) and $\omega_{\mathrm D}=2\omega_0$ (blue). (b) Density plot of $T^{*}(\gamma_0,\omega_{\mathrm D})$ as a function of $\gamma_0$ and $\omega_{\mathrm D}$ (same units as in panel (a)).}
\label{fig:FigTstar}
\end{figure}

\renewcommand{\thesection}{B}
\textit{Appendix B: Details on the behavior of $T^*$}---
The behavior of $T^{*}(\gamma_0,\omega_{\mathrm{D}})$ is obtained
numerically and shown in Fig.~\ref{fig:FigTstar}(a) as a function of
$\gamma_0$ for two representative values of $\omega_{\mathrm D}$.
The critical temperature decreases monotonically with increasing system--bath coupling, thereby enlarging the parameter region where the asymptotic inversion can occur. 
For sufficiently strong coupling, $k_{\mathrm B}T^{*}(\gamma_0,\omega_{\mathrm D})$ becomes much smaller than $\omega_0$, allowing the effect to persist deep in the quantum regime. Conversely, increasing the cut-off frequency raises the critical temperature. This behavior is summarized in Fig.~\ref{fig:FigTstar}(b), which shows that $T^{*}$ decreases with $\gamma_0$ at fixed $\omega_{\mathrm D}$ and increases with $\omega_{\mathrm D}$ at fixed $\gamma_0$. In the strong-coupling limit, this dependence can be obtained analytically, yielding
$
T^{*}(\gamma_0,\omega_{\mathrm D})
\simeq
\frac{\lambda_3}{\pi}
\ln\left(\frac{\omega_{\mathrm D}}{2\lambda_3}\right),
$
for $\gamma_0\to\infty$.

\renewcommand{\thesection}{C}
\textit{Appendix C: Role of initial-slip in the Mpemba effect}---
We now show that the initial-slip term is essential for the emergence of the Mpemba effect. Indeed, omitting the term $\gamma(t)Q(0)$ in the homogeneous Langevin equation (obtained by setting $\xi(t)=0$ in Eq.~\eqref{eq:Langevin}) eliminates the possibility of observing the effect. Since this term enters the dynamics strictly through the initial conditions, it modifies exclusively the homogeneous contribution to the solution.

Taking the Laplace transform of this un-slipped homogeneous equation, one obtains
$
\left[ \lambda^2 + \lambda \widehat{\gamma}(\lambda) + \omega_0^2 \right] \widehat{Q}^{\mathrm{ns}}_{\mathrm{h}}(\lambda) = \left[ \lambda + \widehat{\gamma}(\lambda) \right] Q(0) + \omega_0 P(0),
$
which leads to
$
\widehat{Q}^{\mathrm{ns}}_{\mathrm{h}}(\lambda) = \widehat{G}(\lambda) Q(0) + \omega_0 \widehat{\chi}(\lambda) P(0), 
$
where we have used the response function
$
\widehat{\chi}(\lambda) = {1}/[{\lambda^2 + \lambda \widehat{\gamma}(\lambda) + \omega_0^2}]$
and defined
$
\widehat{G}(\lambda) \equiv \left[ \lambda + \widehat{\gamma}(\lambda) \right] \widehat{\chi}(\lambda) = [{1 - \omega_0^2 \widehat{\chi}(\lambda)}]/{\lambda}.
$
Taking the inverse Laplace transform yields
$Q^{\mathrm{ns}}_{\mathrm{h}}(t) = G(t) Q(0) + \omega_0 \chi(t) P(0),$
where
\begin{equation}
G(t) = 1 - \omega_0^2 \int_0^t \mathrm{d}s  \chi(s) = \omega_0^2 \int_t^\infty \mathrm{d}s  \chi(s) = \omega_0^2 \sum_{i=1}^3 \frac{\chi_i}{\lambda_i} \mathrm{e}^{-\lambda_i t}.
\label{eq:G_integral}
\end{equation}
Here, we used the identity $\int_0^\infty \mathrm{d}s \, \chi(s) = \widehat{\chi}(0) = 1/\omega_0^2$ and the expansion $\chi(t) = \sum_{i=1}^3 \chi_i \mathrm{e}^{-\lambda_i t}$.
The corresponding momentum reads
$
P^{\mathrm{ns}}_{\mathrm{h}}(t) = {\dot{Q}^{\mathrm{ns}}_{\mathrm{h}}(t)}/{\omega_0}
= -\omega_0 \chi(t) Q(0) + \dot{\chi}(t) P(0).
$

We first consider the  moderate-strong interaction where the slowest decay mode $\lambda_3$ is real, that leads to
$G(t)
\simeq
{\omega_0^2}
\chi_3e^{-\lambda_3t}/\lambda_3.$
Consequently, we have
$
Q_{\rm h}^{\rm ns}(t)
\simeq
\omega_0\chi_3e^{-\lambda_3t}
\left[
{\omega_0}Q(0)/{\lambda_3}
+
P(0)
\right]
$
and
$P_{\rm h}^{\rm ns}(t)
\simeq
-\lambda_3\chi_3e^{-\lambda_3t}
\left[
{\omega_0}Q(0)/{\lambda_3}
+
P(0)
\right]$.
The homogeneous position variance therefore reads \footnote{Recall that the oscillator is initially prepared in a thermal state at temperature $T_\alpha$, with a diagonal initial CM satisfying elements $\langle Q^2(0) \rangle_\alpha = \langle P^2(0) \rangle_\alpha = {C_\alpha}/{2}$ and $\left\langle Q(0)P(0) + P(0)Q(0) \right\rangle_\alpha = 0$.}
$\langle \left( Q_{\mathrm{h}}^{\mathrm{ns}}(t) \right)^2 \rangle_\alpha = ({C_\alpha}/{2})
\chi_3^2\,
\left({\omega_0}/{\lambda_3}\right)^2
(\omega_0^2+\lambda_3^2)
e^{-2\lambda_3t}.$
The corresponding expressions for the momentum variance, $\langle \left( P_{\mathrm{h}}^{\mathrm{ns}}(t) \right)^2 \rangle_\alpha$,  and for the symmetrized position--momentum correlator, $\left\langle
\{Q_{\rm h}^{\rm ns}(t),P_{\rm h}^{\rm ns}(t)\}
\right\rangle_\alpha/2$, can be derived analogously (see SM).
Collecting these results, the homogeneous contribution to the CM becomes
$\sigma_{\rm h,\alpha}^{\rm ns}(t)
=
\left({\omega_0}/{\lambda_3}\right)^2
\sigma_{\rm h,\alpha}(t)$, where $\sigma_{\rm h,\alpha}(t)$ is the corresponding homogeneous CM obtained in the presence of the initial-slip term (see main text).
Since the homogeneous CM with and without the initial-slip term are simply proportional to each other, the criteria for the asymptotic inversion derived in the main text remain unchanged. The necessary and sufficient condition thus retains its formal structure, $B_{\mathrm{th}}(T)
<
-({
B^{\rm ns}_{\mathrm{c}}
+
B^{\rm ns}_{\mathrm{h}}
})/{2}$, where \(B_{\mathrm{th}}(T)\) is unchanged (see Appendix A), and
$
B^{\rm ns}_{\alpha}
= ({\omega_0}/{\lambda_3})^2B_{\alpha}
$.
As shown in Appendix A, a necessary requirement for this inequality to hold is
${\cal A}>{\cal B}^{\rm ns}$,
where 
$
{\cal B}^{\rm ns}
=
\lim_{T\rightarrow\infty}
[{
B_{\rm c}^{\rm ns}(T)
+
B_{\rm h}^{\rm ns}(T)
}]/
{2k_{\rm B}T}=
\chi_3^2
{\omega_0}
({\omega_0^2+\lambda_3^2})/{\lambda_3^2}^.
$.
Hence, ${{\cal A}}/{{\cal B}^{\rm ns}}
=
{
\lambda_3\gamma_0\omega_D
}/[
{
(\omega_D-\lambda_3)
(\omega_0^2+\lambda_3^2)]
}=1$, since $\lambda_3\gamma_0\omega_D
=
(\omega_D-\lambda_3)
(\omega_0^2+\lambda_3^2)$.
Therefore, the necessary condition for the occurrence of the Mpemba effect is never satisfied, showing that no Mpemba effect can arise in the absence of the initial-slip term.

We now analyze the Mpemba effect in the UD weak-coupling regime, which is also suitable for comparison with the Born-Markov quantum master equation. As discussed in the main text, for $\gamma_0<2\omega_0$ the two slowest roots form the complex-conjugate pair
$\lambda_{2,3}=\lambda\mp i\eta$, with the corresponding weights satisfying $\chi_2=\chi_3^*$.
In this regime, the long-time behavior of the Green function is
$
G(t)
\simeq\omega_0^2 e^{-\lambda t}\left[({\chi_3}/{\lambda_3})
e^{-i\eta t}+{\rm c.c.}\right].
$
The homogeneous position variance is therefore given by
\begin{equation}\begin{split}
     \langle
Q_{\rm h}^{\rm ns}(t)^2
\rangle_\alpha
&=
\frac{C_\alpha}{2}\omega_0^2 e^{-2\lambda t}\bigg[
\frac{|\chi_3|^2}{|\lambda_3|^2}(\omega_0^2+|\lambda_3|^2)
\\&+\frac{\chi_3^2}{\lambda_3^2}(\omega_0^2+\lambda_3^2) e^{-2i\eta t}
+ {\rm c.c.}\bigg].
\end{split}
\end{equation}
We recall that, in the weak-coupling regime, the thermal contribution is given by
$
\langle \Delta Q(t)\rangle_{\mathrm{th}}
=
B_{\rm th,1}(t)e^{-2\lambda t},
$
where $B_{\rm th,1}(t)$ is defined in Eq.~\eqref{eq:sig}. Using again the relation
$
\lambda_3\gamma_0\omega_D
=
(\omega_D-\lambda_3)(\omega_0^2+\lambda_3^2),
$
we can rewrite the coefficient $S_1(\lambda_3)$ as
$S_1(\lambda_3)=\frac{m (\omega_0^2+\lambda_3^2)}{\lambda_3} F(\lambda_3)$.
Using this identity, the homogeneous and thermal contributions can be combined into the compact form
\begin{equation}
\begin{split}
 &\langle
Q_{\rm h}^{\rm ns}(t)^2
\rangle_\alpha+\langle \Delta Q(t)\rangle_{\mathrm{th}}=
\omega_0\,e^{-2\lambda t}\bigg\{\frac{(\omega_0^2+\lambda_3^2)}{\lambda_3}\cdot\\&
\left[\frac{\omega_0 C_\alpha}{2}-F(\lambda_3)\right]\left[
\frac{|\chi_3|^2}{\lambda}
+\frac{\chi_3^2}{\lambda_3}e^{-2i\eta t}
 \right]\,+{\rm c.c.}\bigg\}.
 \end{split}
 \end{equation}
Similar expressions can be obtained for the momentum variance and for the symmetrized position–momentum correlator (see SM).
These exact expressions provide the starting point for the analysis of the difference between the Frobenius distances. 
By studying numerically the behaviour of
$
\Delta\mathcal{D}(t)=e^{-2\lambda t}\mathcal{F}^{\rm ns}(t),
$
one finds that, for temperatures sufficiently larger than the characteristic scale
$
{\lambda}/{2\pi k_B},
$
the function $\mathcal{F}^{\rm ns}(t)$ remains negative at all times (see SM). Consequently,
$
\Delta\mathcal{D}(t)<0,
$
indicating that the Mpemba effect cannot occur in this regime. 
Extensive numerical investigations in this temperature range confirm that the behaviour described above is generic, and that the initial slip term is essential for the emergence of the Mpemba effect.

We can now briefly discuss the Markovian limit of the weak-coupling dynamics, obtained by assuming
$
k_B T \gg \omega_D \gg \gamma_0,\omega_0 .
$
In this limit, the bath correlation time becomes negligible compared with the characteristic time scales of the system, and one has
$
F(\lambda_3)\simeq k_B T .
$
Moreover, the two slow roots reduce to the standard underdamped Markovian poles, yielding
$
\lambda={\gamma_0}/{2},
\,\,
\eta^2=\omega_0^2-\lambda^2,
\,\,
\chi_3={i}/{2\eta}.
$
Using these expressions, the temperature-dependent prefactor simplifies as
$
{\omega_0}C_\alpha/2-F(\lambda_3)
\simeq
k_B(T_\alpha-T).
$
Therefore, the CM takes the general form
$
\sigma^{\rm ns}_{\alpha}(t)= \sigma(\infty)+\Delta\sigma^{\rm ns}_{\alpha}(t)= \frac{k_{\rm B}T}{\omega_0}+\,\frac{k_{\rm B}(T_{\alpha}-T)}{\omega_0}\,e^{-2\lambda t}\,\mathbb{M}^{\rm ns}(t)
$, where the matrix  $\mathbb{M}^{\rm ns}(t)$ is provided in the SM. 
Therefore, the asymptotic distance is given by
$\mathcal{D}_{\alpha}(t)
=
{k_{\rm B}(T_{\alpha}-T)\,e^{-2\lambda t}}
\sqrt{\tr^2\{y^{\rm ns}_\alpha\}
+2\,\tr\{(y^{\rm ns}_\alpha)^2\}}/{4\omega_0 D_{\infty}^{1/4}}$,
with
$y^{\rm ns}_\alpha=\sigma(\infty)^{-1}\mathbb{M}^{\rm ns}(t)$.
Consequently, the difference between the two asymptotic distances always takes the form
$
    D_{\rm c}(t)-D_{\rm h}(t)=(T_{{\rm c}}-T_{\rm h})\,{\cal G}(t)
$
with \({\cal G}(t)>0\). Since $T_{\rm h} > T_{\rm c}$, it follows that $D_{\rm c}(t) < D_{\rm h}(t)$ at all times, ruling out the Mpemba effect. 
This conclusion is derived directly from the exact solution in the absence of initial-slip and does not rely on any additional master-equation 
approximation.

In the Markovian limit, these findings match those from the standard Born--Markov master equation \cite{Petruccione} (see SM). However, this agreement holds only when omitting the initial-slip term. As shown in the SM, retaining the initial-slip contribution within the exact Caldeira--Leggett dynamics preserves the Mpemba crossing even in the high-temperature regime.

\end{document}


\title{Supplemental Material for \\
``Initial-Slip Dynamics Enables the Quantum Mpemba Effect''}

\author{Maristella Crotti}
\affiliation{\como}
\affiliation{\infnM}
\author{Fabio Cavaliere}
\affiliation{\genova}
\affiliation{\CNR}
\author{Dario Ferraro}
\affiliation{\genova}
\affiliation{\CNR}
\author{Giuliano Benenti}
\affiliation{\como}
\affiliation{\infnM}
\author{Maura Sassetti}
\affiliation{\genova}
\affiliation{\CNR}

\date{\today}

\begin{abstract}
This Supplemental Material provides details about the theoretical framework and mathematical derivations supporting the main text "Initial-Slip Dynamics Enables the Quantum Mpemba Effect". After deriving the Quantum Langevin Equation within the Caldeira-Leggett model, we detail the evaluation of the thermal covariance matrix and the analytical evaluation of Matsubara sums. We then analyze the temperature regime $k_B T < \lambda_3 / 2\pi$, discuss the formal properties of the Frobenius distance 
as an estimator of the distance from equilibrium, and present the theoretical framework for the inverse Mpemba effect. Furthermore, we extend our results to general non-Drude, rational memory kernels, and conclude by proving that the initial time-slip term is indispensable for the emergence of the quantum Mpemba effect. 
\end{abstract}
\maketitle

\tableofcontents

\section{Derivation of the Quantum Langevin Equation}
Our system consists of a quantum harmonic oscillator coupled to an environment modeled as a reservoir of independent bosonic modes \cite{CaldeiraLeggett}. The total Hamiltonian of the system plus reservoir is given by $H = H_{\rm osc} + H_{\rm res} + H_{\rm int}$, where
\begin{align}
H_{\rm osc} &= \frac{p^2}{2m} + \frac{1}{2}m\omega_0^2 x^2, \label{eq:sm_h_osc} \\
H_{\rm res} &= \sum_k \left( \frac{p_k^2}{2m_k} + \frac{1}{2}m_k\omega_k^2 x_k^2 \right), \label{eq:sm_h_res} \\
H_{\rm int} &= -x \sum_k c_k x_k + \frac{1}{2}x^2 \sum_k \frac{c_k^2}{m_k\omega_k^2}. \label{eq:sm_h_int}
\end{align}
Here, $x, p$ are the position and momentum operators of the system oscillator (with mass $m$ and bare frequency $\omega_0$), while $x_k, p_k$ are the position and momentum operators of the $k$-th mode of the reservoir (with mass $m_k$ and frequency $\omega_k$). The constant $c_k$ characterizes the coupling strength between the system and the $k$-th reservoir mode. 
The last term in Eq.~\eqref{eq:sm_h_int} is the standard potential-renormalization counter-term~\cite{CaldeiraLeggett,Ingold02,Weiss2012}. It compensates for the static potential shift induced by the linear interaction, ensuring that the minimum of the potential remains at $x=0$ and preserving the physical frequency $\omega_0$.

To work with dimensionless quantities for the system oscillator, we define the quadratures as
$
Q = \sqrt{m\omega_0}\,x, \,\, P = {p}/{\sqrt{m\omega_0}},
$
satisfying the standard canonical commutation relation $[Q, P] = i$ (setting $\hbar = 1$). In terms of these dimensionless quadratures, the Hamiltonian terms read:
\begin{align}
H_{\rm osc} &= \frac{\omega_0}{2} \left( P^2 + Q^2 \right), \\
H_{\rm int} &= - \frac{Q}{\sqrt{m\omega_0}} \sum_k c_k x_k + \frac{Q^2}{2m\omega_0} \sum_k \frac{c_k^2}{m_k\omega_k^2}.
\end{align}

Using the Heisenberg equations of motion $\dot{A} = i[H, A]$, we obtain the dynamic equations for the system coordinate $Q(t)$ and the bath operators $x_k(t)$:
\begin{align}
\ddot{Q}(t) + \left( \omega_0^2 + \sum_k \frac{c_k^2}{m m_k \omega_k^2} \right) Q(t) &= \sqrt{\frac{\omega_0}{m}} \sum_k c_k x_k(t), \label{eq:sm_eom_Q} \\
\ddot{x}_k(t) + \omega_k^2 x_k(t) &= \frac{c_k}{m_k \sqrt{m\omega_0}} Q(t). \label{eq:sm_eom_xk}
\end{align}

Equation~\eqref{eq:sm_eom_xk} is a second-order linear differential equation driven by the system variable $Q(t)$. Its formal operatorial solution, expressed in terms of the initial conditions at $t=0$, is given by
\begin{equation}
x_k(t) = x_k(0) \cos(\omega_k t) + \frac{p_k(0)}{m_k \omega_k} \sin(\omega_k t) + \frac{c_k}{m_k \omega_k \sqrt{m\omega_0}} \int_0^t d\tau\,\sin[\omega_k (t-\tau)] \, Q(\tau).
\label{eq:sm_xk_sol}
\end{equation}
By substituting the formal solution~\eqref{eq:sm_xk_sol} back into the system equation of motion~\eqref{eq:sm_eom_Q}, we obtain:
\begin{equation}
\ddot{Q}(t) + \omega_0^2 Q(t) + \sum_k \frac{c_k^2}{m m_k \omega_k^2} Q(t) - \sum_k \frac{c_k^2}{m m_k \omega_k} \int_0^t d\tau\,\sin[\omega_k (t-\tau)] \, Q(\tau) = \sqrt{\frac{\omega_0}{m}} \xi(t),
\label{eq:sm_Q_intermediate}
\end{equation}
where we have defined the fluctuating noise operator $\xi(t)$ as
\begin{equation}
\xi(t) = \sum_k c_k \left[ x_k(0) \cos(\omega_k t) + \frac{p_k(0)}{m_k \omega_k} \sin(\omega_k t) \right].
\label{eq:sm_def_xi}
\end{equation}
To rephrase the integral term in Eq.~\eqref{eq:sm_Q_intermediate}, we perform an integration by parts. Noting that 
\begin{equation}
\sin[\omega_k(t-\tau)] = \frac{1}{\omega_k} \frac{\partial}{\partial \tau} \cos[\omega_k(t-\tau)],
\end{equation}
we evaluate:
\begin{equation}
\int_0^t d\tau\,\frac{\sin[\omega_k(t-\tau)]}{\omega_k} \, Q(\tau) = \frac{1}{\omega_k^2} Q(t) - \frac{1}{\omega_k^2} \cos(\omega_k t) Q(0) - \int_0^t d\tau\,\frac{\cos[\omega_k(t-\tau)]}{\omega_k^2} \, \dot{Q}(\tau).
\end{equation}
Inserting this identity back into Eq.~\eqref{eq:sm_Q_intermediate}, the static term $\sum_k \frac{c_k^2}{m m_k \omega_k^2} Q(t)$ originating from the integration by parts precisely cancels out the counter-term contribution. 
Defining the {damping (or memory) kernel} $\gamma(t)$ as
\begin{equation}
\gamma(t) = \frac{1}{m} \sum_k \frac{c_k^2}{m_k \omega_k^2} \cos(\omega_k t),
\label{eq:sm_def_gamma}
\end{equation}
we finally arrive at the {quantum Langevin equation} for the dimensionless quadrature $Q(t)$:
\begin{equation}
\ddot{Q}(t) + \omega_0^2 Q(t) + \int_0^t d\tau\,\gamma(t-\tau)\dot{Q}(\tau) + \gamma(t)Q(0) = \sqrt{\frac{\omega_0}{m}}\xi(t).
\label{eq:sm_Langevin_final}
\end{equation}

\section{Beyond the Drude model}

The results derived in the main text for the Drude bath naturally raise the question of whether this phenomenon relies on the specific form of the Drude spectral density or instead reflects a more general property of dissipative non-Markovian environments. In this Section, we address this issue by
extending the analysis to a broad class of rational memory kernels and identifying which features of the strong-coupling mechanism are universal and which depend on the detailed structure of the bath. For simplicity, we restrict our analysis to the moderate- and strong-coupling regimes.

\subsection{General rational memory kernels}
\label{General kernel}

We consider retarded Ohmic environments with finite memory, whose damping
kernel admits a rational Laplace representation. Specifically, we restrict
our analysis to memory kernels satisfying
$
\gamma(t<0)=0,
\,\,
\lim_{t\to\infty}\gamma(t)=0,
$
together with the low-frequency Ohmic condition $J(\omega\to0^+)\propto\omega$.
Furthermore, we assume spectral densities that remain finite and strictly positive at finite frequencies, thereby excluding transparency windows, namely finite-frequency modes that are completely decoupled from the environment.
Using the relation
$
J(\omega)
=
m\omega\,\Re\hat\gamma(-i\omega),
$
the Ohmic condition implies
$
\hat\gamma(0)>0,
$
where $\hat\gamma(\lambda)$ denotes the Laplace transform of
$\gamma(t)$.

Finally, in order to preserve a finite sum-of-exponentials
representation for both the response function and the bath correlation function, we consider the class of rational damping kernels
$
\hat\gamma(\lambda)
=
\gamma_0
{N(\lambda)}/{D(\lambda)}.
$
The parameter $\gamma_0$ sets the overall coupling strength to the
environment, whereas the polynomials \(N(\lambda)\) and $D(\lambda)$
encode the memory structure of the bath. We choose them as real
polynomials satisfying
$\deg N < \deg D$,
so that $\hat\gamma(\lambda)\to0$ for
$\lambda\to\infty$. Moreover, the condition
$\hat\gamma(0)>0$ requires
\begin{equation}
D(0)\neq0,
\,\,
N(0)\neq0,
\,\,
{N(0)}/{D(0)}>0.
\label{eq:rational_conditions}
\end{equation}
The corresponding response function is
\begin{equation}
\hat\chi(\lambda)
=
\frac{1}
{\lambda^2+\lambda\hat\gamma(\lambda)+\omega_0^2}.
\label{eq:response_general}
\end{equation}
The characteristic roots $\lambda_i$ are determined by the equation
\begin{equation}
Q(-\lambda)
=
(\lambda^2+\omega_0^2)D(-\lambda)
-
\gamma_0\lambda N(-\lambda)
=
0 .
\label{eq:characteristic_general}
\end{equation}
We restrict to the stable, non-degenerate sector, where the roots of
Eq.~\eqref{eq:characteristic_general} are simple and satisfy
$
\Re\lambda_i>0 .
$
Under these assumptions, the response function admits the finite
exponential decomposition
$
\chi(t)
=
\sum_{i=1}^{N_\lambda}
\chi_i e^{-\lambda_i t},
$
where $N_\lambda$ denotes the number of characteristic roots of Eq.~\eqref{eq:characteristic_general}. Since
$\deg N<\deg D$, one has $N_\lambda=\deg D+2$.
The coefficients $\chi_i$ are obtained from the residues of the Laplace-space response function,
$
\hat\chi(\lambda)
=
{D(\lambda)}/
{Q(\lambda)} .
$
The poles of \(\hat\chi(\lambda)\) are located at
\(\lambda=-\lambda_i\), equivalently determined by
\(Q(-\lambda_i)=0\). For simple poles, the corresponding residues are
$
\chi_i
=
-
{D(-\lambda_i)}/
{Q'(-\lambda_i)},
$
where $Q'(\lambda)=\frac{dQ(\lambda)}{d\lambda}$.
Because $Q(\lambda)$ has real coefficients, its roots are either real or occur in complex-conjugate pairs. The same property is inherited by
the coefficients \(\chi_i\). In particular, a real root generates a real
coefficient, while a complex-conjugate pair satisfies
$
(\lambda_i,\lambda_i^*)
\,\,\rightarrow\,\,
(\chi_i,\chi_i^*) .
$
As a result, the response function $\chi(t)$ remains real at all times.

In the strong-coupling regime, the slow characteristic root can be obtained by expanding Eq.~\eqref{eq:characteristic_general} around
$\lambda=0$. Using Eq.~\eqref{eq:rational_conditions}, one finds
\begin{equation}
\lambda_s
=
\frac{\omega_0^2D(0)}
{\gamma_0 N(0)}
+
O(\gamma_0^{-2}).
\label{eq:slow_root_general}
\end{equation}
This root is continuously connected to the solution $\lambda=0$ of the
limiting equation obtained by dividing Eq.~\eqref{eq:characteristic_general}
by $\gamma_0$ and taking the limit $\gamma_0\to\infty$,
$
\lambda N(-\lambda)=0.
$
Since $N(0)\neq0$, the solution \(\lambda=0\) is simple and gives rise to a unique characteristic root of the full problem. Therefore,
$\lambda_s$ is the only root approaching the origin in the
strong-coupling limit.
Moreover, because the spectral density is assumed to be strictly positive
at finite frequencies, $J(\omega)>0$ for $\omega>0$, the equation $N(-\lambda)=0$ has no purely imaginary solutions. Consequently, all remaining roots remain separated from the origin as $\gamma_0\to\infty$. Hence, for sufficiently large $\gamma_0$,
$
\lambda_s<\Re\lambda_i,
\,\,
i\neq s ,
$
and the slow root $\lambda_s$ governs the long-time dynamics.

The bath correlation function retains the same exponential
structure as in the Drude case,
$
L(t)
=
\sum_k
L_k e^{-A_k|t|},$
with
$
\Re (A_k)>0
$ (see next section).
The decay rates $A_k$ are determined by the poles of the memory kernel and by the Matsubara frequencies,
\begin{equation}
\{A_k\}
=
\{A_r^{\rm bath}\}
\cup
\left\{
A_n={2\pi n k_B T}
\right\}_{n\geq1},
\label{eq:Ak_set}
\end{equation}
where the bath poles satisfy $\Re A_r^{\rm bath}>0$.
Since the poles $A_r^{\rm bath}$ depend only on the structure of the environment, they remain finite in the strong-coupling limit
$\gamma_0\to\infty$. Conversely, the slow characteristic root scales
as $\lambda_s\propto\gamma_0^{-1}$, and therefore
$\lambda_s<\Re A_r^{\rm bath},\,\forall r$.
As a consequence, the derivation of the strong-coupling Mpemba
condition follows without substantial modifications. The relevant quantities can be obtained from the Drude expressions by the replacements $\lambda_3\rightarrow\lambda_s, \, \chi_3\rightarrow\chi_s$.

Notice that in the Drude model two temperature regimes were
identified, depending on the relative magnitude of \(k_B T\) and $\lambda_3$. In the present strong-coupling regime,
\(\lambda_s\propto\gamma_0^{-1}\), and therefore, for any finite
temperature, the condition $k_B T>{\lambda_s}/{2\pi}$
is asymptotically satisfied as $\gamma_0\to\infty$. We therefore restrict our analysis to this regime.

The necessary and sufficient condition for the Mpemba effect
retains the same structure as in the Drude case,
$
B_{\rm th} < - \left(B_{\rm c}+B_{\rm h} \right)/2,
\label{inequality}
$
where
\begin{equation}
 B_{\alpha}
 =
 \frac{C_\alpha}{2}\chi_s^2
 \left(\omega_0^2+\lambda_s^2\right)
 \underset{\lambda_s\ll\omega_0}{\simeq}
 \frac{\chi_s^2\omega_0^2}{2}
 \coth\left(\frac{\omega_0}{2k_BT_\alpha}\right),
\end{equation}
and
\begin{equation}
B_{\rm th}
=
-\frac{\omega_0\chi_s^2}{m\lambda_s}
\sum_k
\frac{L_k}
{A_k-\lambda_s}
\underset{\lambda_s\ll A_k}{\simeq}
-
\frac{\omega_0\chi_s^2}{m\lambda_s}
\sum_k
\frac{L_k}{A_k}
=
-\frac{\omega_0\chi_s^2}{m\lambda_s}
\int_0^\infty dt\,L(t)
=
-\frac{\omega_0\chi_s^2}{\beta\lambda_s}
\hat\gamma(0)
=
-\frac{\chi_s^2 k_BT}{\omega_0}
\hat\gamma^2(0).
\label{eq:Bth_general_main}
\end{equation}

Remarkably, the thermal contribution grows quadratically with the effective damping strength, whereas \(B_{\mathrm c}\) and $B_{\mathrm h}$ remain finite for fixed initial temperatures. Consequently, for any finite bath temperature $T>0$, the Mpemba condition is eventually satisfied upon increasing the system--bath coupling.
Therefore, although the detailed dependence of the threshold
temperature on the bath structure is generally model-dependent, the existence of the direct Mpemba effect at sufficiently strong coupling is
universal within the entire class of rational memory kernels considered here. Moreover, for fixed bath temperature $T$, the asymptotic boundary of the Mpemba region in the $(T_{\mathrm c},\Delta T)$ plane follows from
the inequality $
B_{\rm th} < - \left(B_{\rm c}+B_{\rm h} \right)/2
$. In the strong-coupling limit, the corresponding maximum values of $T_{\mathrm c}$ and $\Delta T$ scale as
\begin{equation}
T_{\rm c}^{\rm max}
=
T\left(\frac{\hat\gamma(0)}
{\omega_0}\right)^2,
\,\,\,
\Delta T^{\rm max}
=
2T\left(\frac{\hat\gamma(0)}
{\omega_0}\right)^2 .
\end{equation}

This result shows that increasing dissipation systematically enlarges the
parameter region in which the direct Mpemba effect can be observed. At
leading order, this enhancement is determined solely by the low-frequency
Ohmic response through the single quantity \(\hat\gamma(0)\), whereas the
detailed spectral structure of the environment enters only through
subleading corrections.

\subsection{Bath correlation function beyond Drude}
\label{Lkappa}
We show that the same class of rational memory kernels leads to a finite exponential decomposition of the symmetric bath correlation function.
Since $\hat\gamma(\lambda)$ is a strictly proper rational function and the memory kernel has finite memory, the poles $\lambda_r$ of
$\hat\gamma(\lambda)$, determined by $D(\lambda_r)=0,$ lie in the left half-plane, $\Re\lambda_r<0.$
The symmetric bath correlation function is given by
\begin{equation}
L(t)
=
\int_0^\infty
\frac{d\omega}{\pi}
\,J(\omega)
\coth\!\left(
\frac{\beta\omega}{2}
\right)
\cos(\omega t),
\label{eq:L_definition_general}
\end{equation}
where $J(\omega)=m\omega\,\Re\hat\gamma(-i\omega)$.

To perform the contour integration, we introduce the meromorphic continuation
\begin{equation}
\mathcal J(z)
=
\frac{m z}{2}
\left[
\hat\gamma(-iz)
+
\hat\gamma(iz)
\right],
\label{eq:J_meromorphic}
\end{equation}
which reduces to $J(\omega)$ on the real axis. For $t>0$, the
correlation function can then be expressed as
\begin{equation}
L(t)
=
\int_{-\infty}^{+\infty}
\frac{dz}{2\pi}
\,
\mathcal J(z)
\coth\!\left(
\frac{\beta z}{2}
\right)
e^{izt}.
\label{eq:L_contour_integral}
\end{equation}

The integrand contains two classes of poles. The first class originates
from the poles of the bath response \(\mathcal J(z)\), which are inherited
from the poles \(\lambda_r\) of \(\hat\gamma(\lambda)\). They generate the
decay rates
\begin{equation}
A_r^{\rm bath}
=
-\lambda_r,
\qquad
\Re A_r^{\rm bath}>0 .
\end{equation}
Their total number is \(d=\deg D\), corresponding to the degree of the
polynomial \(D(\lambda)\). The second class arises from the Matsubara
poles of the hyperbolic cotangent,
\begin{equation}
z=-iA_n,
\qquad
A_n=\frac{2\pi n}{\beta},
\qquad
n\geq1 .
\end{equation}

The poles of the hyperbolic cotangent generate the Matsubara decay
rates $A_n$. Closing the contour in the upper (lower) half-plane for $t>0$ ($t<0$), the correlation function is obtained as a sum over
residues. Therefore,
\begin{equation}
L(t)
=
\sum_k L_k e^{-A_k |t|},
\qquad
\Re A_k>0,
\label{eq:L_general_expansion}
\end{equation}
where
\begin{equation}
\{A_k\}
=
\{A_r^{\rm bath}\}_{\rm bath}
\cup
\left\{
A_n=\frac{2\pi n}{\beta}
\right\}_{n\geq1}.
\label{eq:Ak_set}
\end{equation}

The coefficients $L_k$ naturally separate into bath-pole and Matsubara
contributions,
\begin{equation}
L(t)
=
\sum_{r=1}^{d} L_r^{\rm bath}
e^{-A_r^{\rm bath}|t|}
+
\sum_{n=1}^{\infty}
L_n^{M}
e^{-A_n |t|}.
\label{eq:L_split}
\end{equation}
For each pole $z_r=-iA_r^{\rm bath}$ inherited from the bath response,
the corresponding contribution is
\begin{equation}
L_r^{\rm bath}
=
\cot\!\left(
\frac{\beta A_r^{\rm bath}}{2}
\right)
\operatorname*{Res}_{z=-i A_r^{\rm bath}}\mathcal J(z)
=
\frac{m\gamma_0 A_r^{\rm bath}}{2}
\cot\!\left(
\frac{\beta A_r^{\rm bath}}{2}
\right)
\frac{N(-A_r^{\rm bath})}
{D'(-A_r^{\rm bath})}.
\label{eq:Lr_bath}
\end{equation}
The Matsubara contributions arise from the poles of the hyperbolic
cotangent,
\begin{equation}
z_n=-iA_n,
\qquad
A_n=\frac{2\pi n}{\beta},
\qquad
n\geq1 .
\end{equation}
Using $\operatorname*{Res}_{z=-iA_n}
\coth\!\left(\frac{\beta z}{2}\right)
=
\frac{2}{\beta},$
one obtains
$L_n^{M}
=
-m\,k_BT A_n
\left[
\hat\gamma(iA_n)
+
\hat\gamma(-iA_n)
\right].
$

Finally, we note that the bath poles $A_r^{\rm bath}$ are determined
solely by the structure of the environment, whereas the slow root scales
as
$\lambda_s
=
\frac{\omega_0^2D(0)}
{\gamma_0N(0)}
+
O(\gamma_0^{-2}).$
Hence,
$\lambda_s\to0
\qquad
(\gamma_0\to\infty),$
while the bath decay rates remain finite. Therefore, in the
strong-coupling regime,
$\lambda_s < \Re A_r^{\rm bath},
\,\,
\forall r .
\label{eq:lambda_vs_bath}
$
Equation~\eqref{eq:L_general_expansion} provides the generalization of
the Drude exponential decomposition of the bath correlation function. The rates $A_k$ can in general be complex, but always satisfy
$\Re A_k>0$. Complex rates occur in conjugate pairs, ensuring that $L(t)$ remains real.

\subsection{Representative examples}
\label{Examples}
The class of rational damping kernels includes several physically relevant environments commonly employed in the theory of open quantum systems. The simplest example is the standard Drude--Lorentz bath, 
\begin{equation}
\hat\gamma(\lambda)
=
\gamma_0
\frac{\omega_D}{\lambda+\omega_D},
\end{equation}
which corresponds to the spectral density
\begin{equation}
J(\omega)=m\gamma_0
\frac{\omega_D^2\omega}
{\omega^2+\omega_D^2}.
\end{equation}
This is the model analyzed in the main text and describes a bath with a
single memory timescale $\omega_D^{-1}$. A natural extension is provided by multi-Drude environments,
\begin{equation}
\hat\gamma(\lambda)
=
\gamma_0
\sum_{r=1}^{M}
\frac{c_r\omega_r}
{\lambda+\omega_r},
\quad
c_r>0,
\quad
\omega_r\in\mathbb{R}.
\end{equation}
These models describe environments characterized by several memory timescales. Combining the terms over a common denominator gives the
rational form
$\hat\gamma(\lambda)=\gamma_0{N(\lambda)}/{D(\lambda)},$
with $\deg N=M-1$ and $\deg D=M$. Such representations are commonly used to model complex solvents, dielectric environments, and to construct
hierarchical-equation-of-motion approaches, where bath correlation functions are approximated by sums of exponential contributions
\cite{Tanimura1989,Tanimura2020,Weiss2012}.

Another important example is provided by structured resonant
baths described by a Brownian (Lorentzian) spectral density
\cite{Thorwart2004,Groblacher2015,Torre2015}, which appears in molecular
spectroscopy, electron-transfer theory, and vibrational environments
\cite{Leggett1987,Weiss2012,Mukamel1995},
\begin{equation}
J(\omega)
=
m\gamma_0
\frac{\kappa\,\bar\Omega^2\,\omega}
{\left(\bar\Omega^2-\omega^2\right)^2+\kappa^2\omega^2}.
\label{eq:Brownian_J}
\end{equation}
Here, $\gamma_0$ sets the overall damping strength,
$\bar\Omega$ is the resonance frequency, and $\kappa$ determines the width of the resonance. At low frequencies,
$J(\omega)
\sim
m\gamma_0
{\kappa}\omega/\bar\Omega^2$,
showing that the spectral density remains Ohmic. Using
$J(\omega)
=
m\omega\,\Re\hat\gamma(-i\omega),$
the corresponding damping kernel is
\begin{equation}
\hat\gamma(\lambda)
=
\gamma_0
\frac{\lambda+\kappa}
{\lambda^2+\kappa\lambda+\bar\Omega^2},
\label{eq:Brownian_gamma}
\end{equation}
which belongs to the rational class
\(\hat\gamma(\lambda)=\gamma_0N(\lambda)/D(\lambda)\).

More generally, rational kernels naturally arise in finite
pseudomode and reaction-coordinate descriptions of structured
environments. Eliminating a finite number of auxiliary modes typically generates memory kernels with rational Laplace transforms
\cite{Garraway1997,Chin2010,IlesSmith2014}. This provides a broad physical motivation for considering the class of models studied here.

\section{Thermal covariance contribution}

\subsection{Bath correlation function}

In this Section, we derive an exponential expansion of the symmetric part of the bath correlator $\langle \xi(t)\xi(0)\rangle$, denoted by $L(t)$, in terms of the Matsubara frequencies and show how this representation can be used to evaluate the thermal contributions
$\langle Q^2(t) \rangle_{\rm th}$ and
$\langle P^2(t) \rangle_{\rm th}$ presented in the main text.

We start from the definition
\begin{equation}
L(t) =
\int_0^\infty \frac{d\omega}{\pi}\,
J(\omega)
\coth\!\left(\frac{\beta\omega}{2}\right)
\cos(\omega t)
=
m\gamma_0\omega_D^2
\int_{-\infty}^{\infty}
\frac{d\omega}{2\pi}\,
\frac{\omega}{\omega^2+\omega_D^2}
\coth\!\left(\frac{\beta\omega}{2}\right)
\cos(\omega t),
\label{app:symmetric_bath_kernel}
\end{equation}
with $\beta=1/k_BT$. In the second equality, we have substituted the explicit expression of the Drude--Lorentz spectral density $J(\omega)$ and extended the integration domain to the whole real axis.
We then employ the standard partial-fraction expansion of the hyperbolic cotangent,
\begin{equation}
\coth\!\left(\frac{\beta\omega}{2}\right)
=
\frac{2}{\beta\omega}
\left[
1+
2\sum_{n=1}^{\infty}
\frac{\omega^2}{\omega^2+A_n^2}
\right],
\end{equation}
where
$
A_n={2\pi n}/{\beta}
\,\,\, (n\ge1)
$
are the Matsubara frequencies.

Substituting this expansion into Eq.~\eqref{app:symmetric_bath_kernel} and using
$\cos(\omega t)=\tfrac12(e^{i\omega t}+e^{-i\omega t})$, together with the symmetry of the integrand, we obtain
\begin{equation}
L(t)
=
\frac{m\gamma_0\omega_D^2}{\beta}
\int_{-\infty}^{\infty}
\frac{d\omega}{\pi}\,
\frac{1}{\omega^2+\omega_D^2}
\left[
1+
2\sum_{n=1}^{\infty}
\frac{\omega^2}{\omega^2+A_n^2}
\right]
e^{i\omega t}.
\label{app:1symmetric_bath_kernel}
\end{equation}
The integral above can be performed in the complex plane by closing the contour in the upper (lower) half-plane for $t>0$ ($t<0$). Here, we always assume to have simple poles, thereby excluding cases where $\omega_D = A_n$ for any $n \ge 1$. Under this assumption, applying the residue theorem yields
\begin{equation} 
L(t)=\frac{m\gamma_0\omega_D^2}{2}\left[\cot\!\left(\frac{\beta\omega_D}{2}\right)e^{-\omega_D|t|}+\frac{4}{\beta}\sum_{n=1}^\infty\frac{A_n e^{-A_n|t|}}{A_n^2-\omega_D^2} \right],
\label{app:symmetric_bath_kernel_result} 
\end{equation}
that can be recast in the compact form
\begin{equation}
\label{app:Ls_sum}
L(t) = \sum_{n=0}^{\infty} L_n e^{-A_n |t|},
\end{equation}
where the expansion coefficients $L_n$ and the decay rates $A_n$ are defined as:
\begin{align}
&L_0 = \frac{m \gamma_0 \omega_D^2}{2} \cot\left(\frac{\beta \omega_D}{2}\right),  \quad A_0 = \omega_D, \nonumber \\
&L_{n \ge 1} = \frac{m \gamma_0 \omega_D^2}{\beta} \frac{2 A_n}{A_n^2 - \omega_D^2},  \quad A_{n \ge 1} = \frac{2 \pi n}{\beta}.
\label{app:L0_definition}
\end{align}

\subsection{Thermal contribution to covariance}
We now show in details how from this form of $L(t)$ we can evaluate the thermal contiributions
\begin{equation}
\begin{split}
\langle Q^2(t)\rangle_{\mathrm{th}}
&=
\frac{\omega_0}{m}
\int_0^t d\tau_2
\int_0^t d\tau_1
\chi(\tau_1)\chi(\tau_2)
L(\tau_2-\tau_1),\nonumber\\
\langle P^2(t)\rangle_{\mathrm{th}}
&=
\frac{1}{m\omega_0}
\int_0^t d\tau_2
\int_0^t d\tau_1
\dot\chi(\tau_1)\dot\chi(\tau_2)
L(\tau_2-\tau_1).
\label{eq:thermal_general}
\end{split}
\end{equation}
The integration domain can be split as
\begin{eqnarray}
\langle Q^2(t)\rangle_{\mathrm{th}}
&=&
\frac{2\omega_0}{m}
\int_0^t d\tau_2
\int_0^{\tau_2} d\tau_1\,
\chi(\tau_1)\chi(\tau_2)
L(\tau_2-\tau_1)\nonumber\\
\langle P^2(t)\rangle_{\mathrm{th}}
&=&
\frac{2}{m\omega_0}
\int_0^t d\tau_2
\int_0^{\tau_2} d\tau_1\,
\dot\chi(\tau_1)\dot\chi(\tau_2)
L(\tau_2-\tau_1).
\label{eq:thermal_triangular}
\end{eqnarray}
Substituting Eq.~\eqref{app:Ls_sum} one obtains
\begin{eqnarray}
\langle Q^2(t)\rangle_{\mathrm{th}}
&=&
\sum_{i=1}^{3}
\sum_{j=1}^{3}
\sum_{n=0}^{\infty}
\chi_i\chi_j\,L_n\,
I_{ijn}{(t)},\nonumber\\
\langle P^2(t)\rangle_{\mathrm{th}}
&=&
\frac{1}{\omega_0^2}
\sum_{i=1}^{3}
\sum_{j=1}^{3}
\sum_{n=0}^{\infty}
(\lambda_i\chi_i)
(\lambda_j\chi_j)
L_n\,
I_{ijn}{(t)},
\label{eq:thermal_sum}
\end{eqnarray}
with the integral \(I_{ijn}{(t)}\) defined as
\begin{equation}
I_{ijn}{(t)}
=\frac{2\omega_0}{m}
\int_0^t d\tau_2
\int_0^{\tau_2} d\tau_1\,
e^{-\lambda_i\tau_2}
e^{-\lambda_j\tau_1}
e^{-A_n(\tau_2-\tau_1)}.
\label{eq:Iijn_definition}
\end{equation}
Performing integrations we have
\begin{equation} 
I_{ijn}{(t)} = \frac{2\omega_0}{m} \frac{1}{A_n-\lambda_j} \left[ \frac{1}{\lambda_i+\lambda_j} \left( 1-e^{-(\lambda_i+\lambda_j)t} \right) + \frac{1}{\lambda_i+A_n} \left( e^{-(\lambda_i+A_n)t}-1 \right) \right]. 
\label{eq:Iijn_exact} 
\end{equation} 
As we can see $I_{ijn}{(t)} $ can be separated into  stationary and transient contributions
\begin{equation} I_{ijn}{(t)} = I_{ijn}{(\infty)} + \Delta I_{ijn}{(t)}, \label{eq:Iijn_split} \end{equation}
with
\begin{eqnarray} I_{ijn}{(\infty)} &=& \frac{2\omega_0}{m} \frac{1} {(\lambda_i+\lambda_j)(\lambda_i+A_n)},\nonumber\\
&&\nonumber\\
\Delta I_{ijn}{(t)}=\Delta I^{(1)}_{ijn}{(t)}+\Delta I^{(2)}_{ijn}{(t)} &=& \frac{2\omega_0}{m}\frac{1}{A_n-\lambda_j} \left[ \frac{e^{-(\lambda_i+A_n)t}} {\lambda_i+A_n} - \frac{e^{-(\lambda_i+\lambda_j)t}} {\lambda_i+\lambda_j} \right].
\label{eq:DeltaIijn} 
\end{eqnarray} 
The corresponding thermal variances therefore are
\begin{eqnarray} 
\langle Q^2(t)\rangle_{\mathrm{th}} &=& \langle Q^2(\infty)\rangle + \langle \Delta Q^{(1)}(t)\rangle_{\mathrm{th}}+\langle \Delta Q^{(2)}(t)\rangle_{\mathrm{th}},\nonumber\\
\langle P^2(t)\rangle_{\mathrm{th}} &=& \langle P^2(\infty)\rangle + \langle \Delta P^{(1)}(t)\rangle_{\mathrm{th}}+\langle \Delta P^{(2)}(t)\rangle_{\mathrm{th}}, 
\label{eq:X2_split} 
\end{eqnarray}
with equilibrium stationary values 
\begin{eqnarray} 
\langle Q^2(\infty)\rangle&=&\sum_{i=1}^{3} \sum_{j=1}^{3} \sum_{n=0}^{\infty} \chi_i \chi_j L_n \, I_{ijn}{(\infty)},\nonumber\\
\langle P^2(\infty)\rangle&=& \frac{1}{\omega_0^2} \sum_{i=1}^{3} \sum_{j=1}^{3} \sum_{n=0}^{\infty} (\lambda_i\chi_i) (\lambda_j\chi_j) L_n \, I_{ijn}{(\infty)}
\label{eq:DeltaX2_thermal} 
\end{eqnarray} 
and transient corrections
\begin{eqnarray} 
\langle \Delta Q^{(1,2)}(t)\rangle_{\mathrm{th}} &=&\sum_{i=1}^{3} \sum_{j=1}^{3} \sum_{n=0}^{\infty} \chi_i \chi_j L_n \, \Delta I^{(1,2)}_{ijn}{(t)},\nonumber\\
\langle \Delta P^{(1,2)}(t)\rangle_{\mathrm{th}} &=& \frac{1}{\omega_0^2} \sum_{i=1}^{3} \sum_{j=1}^{3} \sum_{n=0}^{\infty} (\lambda_i\chi_i) (\lambda_j\chi_j) L_n \, \Delta I^{(1,2)}_{ijn}{(t)}.
\label{eq:DeltaX2_thermal} 
\end{eqnarray} 

\section{Details on the Frobenius distance}
\label{Frobenius}
In the main text, we quantify the distance between the reduced density matrix
$\rho_{\alpha}(t)$ of the oscillator at time $t$ and its asymptotic state $\rho(\infty)$ by means of the Frobenius distance. In this Section, we provide more technical details underlying the corresponding calculations. 

We recall that the Frobenius distance is defined as~\cite{Serafini2017}
\begin{equation}
\mathcal{D}_\alpha(t)
=
\sqrt{\tr\!\left\{
\left[
\rho_\alpha(t)-\rho(\infty)
\right]^2
\right\}}\,.
\end{equation}
For Gaussian states with vanishing first moments,
$\mathcal{D}_\alpha(t)$ can be expressed solely in terms of the covariance matrices as~\cite{Serafini2017}
\begin{equation}
\mathcal{D}_{\alpha}(t)
=
\frac{1}{D_{\infty}^{1/4}}
\sqrt{
1+
\frac{1}
{\sqrt{\det\!\left[\mathbb{I}+y_{\alpha}(t)\right]}}
-
\frac{2}
{\sqrt{\det\!\left[\mathbb{I}+\frac{y_{\alpha}(t)}{2}\right]}}
}\,,
\label{eq:Frobenius1}
\end{equation}
where
$\mathbb{I}=\mathrm{diag}\{1,1\}$,
$D_{\infty}=\det\sigma(\infty)$, and
$
y_{\alpha}(t)
=
\sigma(\infty)^{-1}
\Delta\sigma_{\alpha}(t).
$
Here, from now on, we will focus on both the overdamped (OD) moderate-- and underdamped (UD) strong--coupling regimes.\\

At long times, $y_\alpha(t)$ takes the form
\begin{equation}
y_\alpha(t)
=
\sum_{n=1}^{\infty}
e^{-(\lambda_3+A_n)t}
Q_n\mathbb{M}_n
+
e^{-2\lambda_3t}
\left(B_\alpha+B_{\mathrm{th}}\right)\mathbb{M}_0,
\end{equation}
where sub--leading exponential contributions $\propto e^{-\lambda_{1,2}t}$ have been neglected. Also notice that summations in $y_{\alpha}(t)$ start from $n=1$ since the term with $n=0$ yields a sub--leading term $\propto e^{-(\lambda_3+\omega_{\mathrm{D}})t}$ which again can be neglected in the long time limit. Finally, the $2\times 2$ matrices $\mathbb{M}_0$ and $\mathbb{M}_n$ are defined as
\begin{equation}
\mathbb{M}_0=
\begin{pmatrix}
\frac{1}{\langle Q^2(\infty)\rangle} &
-\frac{\lambda_3}{\omega_0\langle Q^2(\infty)\rangle}\\[1.5ex]
-\frac{\lambda_3}{\omega_0\langle P^2(\infty)\rangle} &
\frac{\lambda_3^2}{\omega_0^2\langle P^2(\infty)\rangle}
\end{pmatrix},
\qquad
\mathbb{M}_n=
\begin{pmatrix}
\frac{1}{\langle Q^2(\infty)\rangle} &
-\frac{\lambda_3+A_n}{2\omega_0\langle Q^2(\infty)\rangle}\\[1.5ex]
-\frac{\lambda_3+A_n}{2\omega_0\langle P^2(\infty)\rangle} &
\frac{\lambda_3A_n}{\omega_0^2\langle P^2(\infty)\rangle}
\end{pmatrix}.
\end{equation}
By construction,
$\mathcal{D}_\alpha(t)\ge0$
and
$\mathcal{D}_\alpha(t)\to0$
as
$t\to\infty$.

\subsection{Leading asymptotics}
\label{app:leadas}
We now show that the asymptotic behavior of
$\Delta\mathcal{D}(t)$ is completely determined by the two contributions
$\tr^2\{y_{\alpha}(t)\}$ and
$\tr\{y_{\alpha}^2(t)\}$,
while all remaining terms decay faster and therefore provide only subleading corrections.
Consequently, in the asymptotic regime,
\begin{equation}
\mathcal{D}_{\alpha}(t)
\approx
\frac{1}{4D_{\infty}^{1/4}}
\sqrt{
\tr^2\{y_{\alpha}(t)\}
+
2\,\tr\{y_{\alpha}^2(t)\}
}\,.
\label{Dtraccia}
\end{equation}

Using the identities
\begin{equation}
\det\!\left[\mathbb{I}+y_{\alpha}(t)\right]
=
1+\tr\{y_{\alpha}(t)\}
+\det y_{\alpha}(t),
\qquad
\det y_{\alpha}(t)
=
\frac12
\left[
\tr^2\{y_{\alpha}(t)\}
-
\tr\{y_{\alpha}^2(t)\}
\right],
\label{eq:rel1}
\end{equation}
together with
$
\tr\!\left[{y_{\alpha}(t)}/{2}\right]
=
\tr\{y_{\alpha}(t)\}/2
$ and $
\tr\!\left[\left({y_{\alpha}(t)}/{2}\right)^2\right]
=
\tr\{y_{\alpha}^2(t)\}/4,
$
the determinant appearing in Eq.~\eqref{eq:Frobenius1} 
can be expressed entirely in terms of
$\tr\{y_{\alpha}(t)\}$ and
$\tr\{y_{\alpha}^2(t)\}$.
Therefore, the late-time expansion of $\mathcal D_\alpha(t)$ 
reduces to an expansion in powers of these two quantities.

A straightforward expansion shows that the constant term and all contributions linear in
$\tr\{y_{\alpha}(t)\}$ cancel identically. The remaining terms therefore consist exclusively of monomials of the form
$
\left[\tr y_{\alpha}(t)\right]^{p_1}
\left[\tr y_{\alpha}^2(t)\right]^{p_2},
$
with either
\begin{enumerate}
\item $p_2=0$ and $p_1\ge2$;
\item $p_1=0$ and $p_2\ge1$;
\item $p_1\ge1$ and $p_2\ge1$.
\end{enumerate}
Our goal is to show that the leading asymptotic behavior of
$\Delta\mathcal{D}(t)$ 
is entirely determined by the terms
$\tr^2\{y_{\alpha}(t)\}$ and
$\tr\{y_{\alpha}^2(t)\}$.
To this end, we first establish the general structure of the traces
$\tr\{y_{\alpha}(t)\}$ and
$\tr\{y_{\alpha}^2(t)\}$.
Using the explicit form of $y_\alpha$ reported above, one finds
\begin{eqnarray}
\tr{y_{\alpha}(t)}
&=&
Y_0(\alpha)e^{-2\lambda_3t}
+\sum_{n=1}^{\infty}
Y_{n}e^{-(\lambda_3+A_n)t},
\label{eq:try}
\\
\tr{y_{\alpha}^2(t)}
&=&
\bar{Y}_0(\alpha)e^{-4\lambda_3 t}
+\sum_{n=1}^{\infty}
\bar{Y}_n^{(1)}(\alpha)e^{-(3\lambda_3+A_n)t}
+\sum_{n=1}^{\infty}\sum_{n'=1}^{\infty}
\bar{Y}_{n,n'}^{(2)}
e^{-\left(2\lambda_3+A_n+A_{n'}\right)t},
\label{eq:try2}
\end{eqnarray}
where
\begin{equation}
Y_0(\alpha)
=
(B_\alpha+B_{\mathrm{th}})
\tr\mathbb{M}_0,
\qquad
Y_n
=
Q_n\tr\mathbb{M}_n,
\label{eq:coefs1}
\end{equation}
\begin{equation}
\bar{Y}_0(\alpha)
=
Y_0^2(\alpha),
\qquad
\bar{Y}_n^{(1)}(\alpha)
=
2(B_\alpha+B_{\mathrm{th}})Q_n
\tr\!\left(\mathbb{M}_0\mathbb{M}_n\right),
\qquad
\bar{Y}_{n,n'}^{(2)}
=
Q_nQ_{n'}
\tr\!\left(\mathbb{M}_n\mathbb{M}_{n'}\right),
\label{eq:coefs2}
\end{equation}
with $n,n'\ge1$.
The expression for $\bar{Y}_0(\alpha)$ follows from the identity
$\tr(\mathbb{M}_0^2)=\tr^2(\mathbb{M}_0)$, while the coefficient
$\bar{Y}_n^{(1)}(\alpha)$ exploits the cyclic invariance of the trace,
$\tr(\mathbb{M}_0\mathbb{M}_n)=\tr(\mathbb{M}_n\mathbb{M}_0)$.
Finally, direct evaluation shows that all traces appearing in
Eqs.~\eqref{eq:try} and \eqref{eq:try2} are strictly positive.

We now distinguish between the two temperature regimes:
\begin{itemize}
\item $k_{\mathrm{B}}T>\lambda_3/2\pi$:
Since
$
A_n={2\pi n}/{\beta}> \lambda_3$ for $
n\ge1,
$
the slowest exponential contribution to
$\tr\{y_{\alpha}(t)\}$ is
$Y_0(\alpha)e^{-2\lambda_3t}$,
whereas the leading contribution to
$\tr\{y_{\alpha}^2(t)\}$ is
$\bar{Y}_0(\alpha)e^{-4\lambda_3t}$.
Consequently, a generic term in the asymptotic expansion has the form
\begin{equation}
\left[\tr^{p_1}y_{\alpha}(t)\right]
\left[\tr^{p_2}y_{\alpha}^2(t)\right]
=
Y_0(\alpha)^{p_1+2p_2}
e^{-4\lambda_3\left(p_2+\frac{p_1}{2}\right)t}.
\end{equation}

From the constraints discussed above [cases (1)--(3)], one always has
$
p_2+{p_1}/{2}\ge1.
$
The minimum value is attained only for
$(p_1,p_2)=(2,0)$
and
$(p_1,p_2)=(0,1)$,
which correspond to the contributions
$\tr^2\{y_{\alpha}(t)\}$
and
$\tr\{y_{\alpha}^2(t)\}$,
respectively.
All remaining terms decay faster and therefore constitute subleading corrections.
Hence, in the regime
$k_{\mathrm{B}}T>\lambda_3/(2\pi)$,
the asymptotic behavior is entirely determined by
\begin{equation}
\tr^2\{y_{\alpha}(t)\}
\simeq
Y_0^2(\alpha)e^{-4\lambda_3t},
\qquad
\tr\{y_{\alpha}^2(t)\}
\simeq
\bar{Y}_0(\alpha)e^{-4\lambda_3t}.
\end{equation}

\item $k_{\mathrm{B}}T<\lambda_3/2\pi$:

In this regime, $\lambda_3>A_1$, and the identification of the leading contributions requires a more careful analysis. We explicitly consider the family of terms
$\tr^{p_1}\{y_{\alpha}(t)\}$ with $p_2=0$ and $p_1\ge2$, proving that only the case $p_1=2$ contributes to the leading asymptotic behavior, whereas all terms with $p_1>2$ are subleading. The analysis for the remaining two families of terms proceeds analogously and is therefore omitted.

We begin with the case $p_1=2$, for which
\begin{equation}
\tr^2y_{\alpha}(t)
=
Y_0^2(\alpha)e^{-4\lambda_3t}
+
2Y_0(\alpha)\sum_{n=1}^{\infty}
Y_ne^{-(3\lambda_3+A_n)t}
+
\sum_{n=1}^{\infty}\sum_{n'=1}^{\infty}
Y_nY_{n'}
e^{-(2\lambda_3+A_n+A_{n'})t}.
\end{equation}
The slowest decaying contribution is
$Y_1^2e^{-2(\lambda_3+A_1)t}$.
However, this term is independent of the initial state and therefore cancels when considering differences of Frobenius distances,
$\Delta\mathcal D(t)$,
between different initial conditions.
Although this contribution is the slowest decaying one, it does not affect the asymptotic behavior of these differences. Consequently, the relevant asymptotic contribution is the slowest term depending explicitly on the initial state, namely
$2Y_0(\alpha)Y_1e^{-(3\lambda_3+A_1)t}$.
Keeping only the leading and the leading $\alpha$-dependent contributions, one finds
\begin{equation}
\tr^2y_{\alpha}(t)
\approx
Y_1^2e^{-2(\lambda_3+A_1)t}
+
2Y_0(\alpha)Y_1
e^{-(3\lambda_3+A_1)t}.
\label{eq:leading}
\end{equation}

We now consider the general case $p\ge3$. To prove that no higher-order power contributes to the asymptotic behavior, it is sufficient to estimate the smallest possible decay exponent among all terms appearing in the multinomial expansion,
\begin{equation}
\left[
Y_0(\alpha)e^{-2\lambda_3t}
+
\sum_{n=1}^{\infty}
Y_ne^{-(\lambda_3+A_n)t}
\right]^p
=
\sum_{m=1}^{p}
\mathcal S_m(t),
\end{equation}
where $\mathcal S_m(t)$ denotes the sum of all distinct products involving exactly $m$ different terms inside the brackets.

An arbitrary term in the multinomial expansion of $\mathcal S_m(t)$ can be written as
\begin{equation}
e^{-\Gamma t}
\prod_{i=1}^{m}
Y_{\mu_i}^{k_i},
\label{eq:bigmess}
\end{equation}
where
$k_i\ge1$,
$\sum_{i=1}^{m}k_i=p$,
and
$0\le\mu_1<\mu_2<\cdots<\mu_m$.
For convenience, we introduce the notation
$Y_0\equiv Y_0(\alpha)$,
$\Lambda_0\equiv2\lambda_3$,
and
$\Lambda_{n\ge1}\equiv\lambda_3+A_n$.
The decay exponent of the generic contribution is
$
\Gamma=\sum_{i=1}^{m}k_i\Lambda_{\mu_i}.
$

Therefore, identifying the leading asymptotic contribution amounts to minimizing $\Gamma$ over all admissible choices of the multiplicities $k_i$ and indices $\mu_i$. Rather than solving this minimization problem exactly, it is sufficient to derive a lower bound on $\Gamma$.

Let $q$ denote the multiplicity of the factor $Y_0$ in Eq.~\eqref{eq:bigmess}. Then
\begin{equation}
\Gamma
=
2q\lambda_3
+
\sum_{\mu_i>0}
k_i(\lambda_3+A_{\mu_i})
=
(p+q)\lambda_3
+
R,
\end{equation}
where
$R=\sum_{\mu_i>0}k_iA_{\mu_i},$
and we used the fact that 
$
\sum_{\mu_i>0}
k_i=p-q,
$
since $q$ factors correspond to $Y_0$.

Since $A_n=nA_1$, we have
$
R
=
A_1
\sum_{\mu_i>0}
k_i\mu_i.
$
Moreover, every index satisfies $\mu_i\ge1$, and therefore
$
\sum_{\mu_i>0}
k_i\mu_i
\ge
\sum_{\mu_i>0}
k_i=p-q.
$
Hence
$
R
\ge
(p-q)A_1,
$
which yields the lower bound
$
\Gamma
\ge
(p+q)\lambda_3
+
(p-q)A_1.
$
Because $p\ge3$, it follows that
$
\Gamma
\ge
(3+q)\lambda_3
+
(3-q)A_1.
$

Finally, comparing this lower bound with the decay exponent of the leading $\alpha$-dependent contribution in Eq.~\eqref{eq:leading}, namely
$3\lambda_3+A_1$, we obtain
$
(3+q)\lambda_3+(3-q)A_1
>
3\lambda_3+A_1,
$
or equivalently,
$ 
(\lambda_3-A_1)q+2A_1>0,
$ 
which is always satisfied since $\lambda_3>A_1$. Therefore every contribution with $p\ge3$ decays faster than the leading $\alpha$-dependent term. Hence no power
$\tr^{p}\{y_\alpha(t)\}$ with $p\ge3$ can modify the asymptotic scaling determined by Eq.~\eqref{eq:leading}, completing the proof.
\end{itemize} 
A completely analogous analysis can be carried out for $\tr\{y_{\alpha}^2(t)\}$. The corresponding leading and subleading asymptotic contributions are \begin{equation} \tr\{y_{\alpha}^2(t)\} \approx \begin{cases} \bar{Y}_0(\alpha)e^{-4\lambda_3t}, & k_{\mathrm{B}}T>\dfrac{\lambda_3}{2\pi}, \\[2mm] \bar{Y}_{1,1}^{(2)}e^{-2(\lambda_3+A_1)t} + \bar{Y}_1^{(1)}(\alpha)e^{-(3\lambda_3+A_1)t}, & k_{\mathrm{B}}T<\dfrac{\lambda_3}{2\pi}, \end{cases} \label{eq:try2as} \end{equation} while all higher-order contributions $\tr^{p}\{y_{\alpha}^2(t)\}$ with $p>1$, as well as mixed terms of the form $\left[\tr^{p_1}\{y_{\alpha}(t)\}\right] \left[\tr^{p_2}\{y_{\alpha}^2(t)\}\right]$ with $p_1,p_2\ge1$, are readily shown to decay faster than the terms reported in Eq.~\eqref{eq:try2as}. 


\section{Details on the Frobenius distance at $t=0$}

\subsection{Ordering of the initial Frobenius distances}
\label{app:initial_distance}

In the discussion of the direct Mpemba effect, the labels ``hot'' and
``cold'' refer to the temperatures of the initial thermal preparations,
with $T_{\rm h}>T_{\rm c}>T$, where $T$ is the temperature of the
environment. In addition, the hot preparation must initially be farther
from the asymptotic state than the cold one. Namely,
\begin{equation}
{\cal D}_{\rm h}(0)>{\cal D}_{\rm c}(0).
\label{eq:initial_ordering}
\end{equation}
Since at finite system--bath coupling the asymptotic reduced state is
not, in general, the Gibbs state of the bare oscillator, the temperature
ordering $T_{\rm h}>T_{\rm c}$ does not automatically imply an analogous
ordering of the initial distances from the asymptotic state. It is
therefore necessary to examine explicitly the condition
${\cal D}_{\rm h}(0)>{\cal D}_{\rm c}(0)$. We now investigate this point.
We start from the Frobenius distance introduced in the main text and in Sec.~\ref{Frobenius}:
\begin{equation}
\mathcal{D}_{\alpha}(t)
=
\frac{1}{D_{\infty}^{1/4}}
\sqrt{
1+
\frac{1}
{\sqrt{\det\!\left[\mathbb{I}+y_{\alpha}(t)\right]}}
-
\frac{2}
{\sqrt{\det\!\left[\mathbb{I}+\frac{y_{\alpha}(t)}{2}\right]}}
}\,,
\label{eq:Frobenius_initial_app}
\end{equation}
where
\begin{equation}
y_{\alpha}(t)
=
\sigma(\infty)^{-1}
\Delta\sigma_{\alpha}(t),
\qquad
\Delta\sigma_{\alpha}(t)
=
\sigma_{\alpha}(t)-\sigma(\infty),
\qquad
D_\infty=\det\sigma(\infty).
\end{equation}

For the thermal initial preparation considered here, the covariance
matrix at $t=0$ is isotropic,
\begin{equation}
\sigma_\alpha(0)=\frac{1}{2}C_\alpha\,\mathbb{I},
\qquad
C_\alpha=\coth\left(\frac{\omega_0}{2k_{\rm B}T_\alpha}\right),
\qquad
\alpha={\rm c},{\rm h}.
\label{eq:nu_initial}
\end{equation}
Since $T_{\rm h}>T_{\rm c}$, one has
$C_{\rm h}>C_{\rm c}$. The stationary covariance matrix is diagonal and
can be written as

\begin{equation}
\sigma(\infty)
=
\begin{pmatrix}
Q_\infty & 0\\
0 & P_\infty
\end{pmatrix},
\qquad
D_\infty=Q_\infty P_\infty,
\label{eq:sigma_infty_QP}
\end{equation}
with the stationary variances  expressed, through the
fluctuation--dissipation theorem, as \cite{Weiss2012}

\begin{align}
Q_\infty
=
\frac{\omega_0}{\pi}
\int_0^\infty d\omega\,
{\rm Im}\,\tilde\chi(\omega)
\coth\left(\frac{\omega}{2k_{\rm B}T}\right),
\qquad
P_\infty
=
\frac{1}{\pi\omega_0}
\int_0^\infty d\omega\,
\omega^2{\rm Im}\,\tilde\chi(\omega)
\coth\left(\frac{\omega}{2k_{\rm B}T}\right).
\label{eq:QP_FDT}
\end{align}

Here, $\tilde\chi(\omega)$ is the Fourier transform of the response function $\chi(t)$ defined in the main part. For the Drude model it is 
\begin{equation}
\tilde\chi(\omega)
=
\frac{1}{
\omega_0^2-\omega^2
-i\omega
\displaystyle\frac{\gamma_0\omega_D}
{\omega_D-i\omega}
}.
\label{eq:chi_Drude}
\end{equation}

At the initial time $
\Delta\sigma_\alpha(0)
=
\frac{C_\alpha}{2}\mathbb{I}
-
\sigma(\infty)$ 
and therefore

\begin{eqnarray}
\det\left[
\mathbb{I}+y_\alpha(0)
\right]
=
\frac{C_\alpha^2}{4D_\infty},
\qquad \det\left[
\mathbb{I}+\frac{y_\alpha(0)}{2}
\right]=
\frac{
\left(Q_\infty+\frac{C_\alpha}{2}\right)
\left(P_\infty+\frac{C_\alpha}{2}\right)
}{
4D_\infty
}.
\label{eq:det2_initial}
\end{eqnarray}

Substituting Eq~\eqref{eq:det2_initial} into Eq.~\eqref{eq:Frobenius_initial_app},
we obtain

\begin{equation}
{\cal D}_\alpha^2(0)
=
\frac{1}{\sqrt{D_\infty}}
+
\frac{2}{C_\alpha}
-
\frac{4}{
\sqrt{
\left(Q_\infty+\frac{C_\alpha}{2}\right)
\left(P_\infty+\frac{C_\alpha}{2}\right)
}
}.
\label{eq:D_initial_exact}
\end{equation}

We now study the conditions on $T_{\rm c}$ and $T_{\rm h}$   for  an initial standard preparation of the  direct Mpemba effect with ${\cal D}^2_{\rm h}(0)>{\cal D}^2_{\rm c}(0)$.

This
question can be addressed by studying the dependence of the initial
distance directly on the thermal parameter $C$.
Indeed, at fixed bath temperature $T$, the stationary quantities
$Q_\infty$, $P_\infty$, and $D_\infty$ are fixed. We can therefore
regard the initial Frobenius distance as a function of the single
thermal parameter $C$ and define
\begin{equation}
F_T(C)
\equiv
{\cal D}^2(0;C)
=
\frac{1}{\sqrt{D_\infty}}
+
\frac{2}{C}
-
\frac{4}{
\sqrt{
\left(Q_\infty+\frac{C}{2}\right)
\left(P_\infty+\frac{C}{2}\right)
}
}.
\label{eq:FT_definition}
\end{equation}
By construction, $F_T(C)\geq0$. For a thermal preparation at finite
temperature, $C=
\coth\left(\frac{\omega_0}{2k_{\rm B}T_\alpha}\right)>1$, 
with $C=1$ at zero temperature. We therefore consider in
the end only values in the physical domain $C>1$.
We now study the monotonicity of $F_T(C)$. Differentiating
Eq.~\eqref{eq:FT_definition}, one finds
\begin{equation}
\frac{dF_T(C)}{dC}
=
\frac{2}{C^2}
\left\{
\frac{
C^2\left(C+Q_\infty+P_\infty\right)
}{
2
\left[
\left(Q_\infty+\frac{C}{2}\right)
\left(P_\infty+\frac{C}{2}\right)
\right]^{3/2}
}
-1
\right\}.
\label{eq:dFT_dC_factorized}
\end{equation}

To determine the sign of the derivative, we note that the quantity
appearing as the first term inside the curly brackets is a strictly
increasing function of $C$, using $Q_\infty>0$ and $P_\infty>0$.
Moreover,
\begin{equation}
\lim_{C\rightarrow0}
\frac{
C^2\left(C+Q_\infty+P_\infty\right)
}{
2
\left[
\left(Q_\infty+\frac{C}{2}\right)
\left(P_\infty+\frac{C}{2}\right)
\right]^{3/2}
}
=0,
\qquad
\lim_{C\rightarrow\infty}
\frac{
C^2\left(C+Q_\infty+P_\infty\right)
}{
2
\left[
\left(Q_\infty+\frac{C}{2}\right)
\left(P_\infty+\frac{C}{2}\right)
\right]^{3/2}
}
=4.
\end{equation}
It follows that the quantity in curly brackets in
Eq.~\eqref{eq:dFT_dC_factorized} vanishes once and only once. We
denote the corresponding value of $C$ by $C_{\min}(T)$. Therefore,
$F_T(C)$ possesses a single minimum at $C=C_{\min}(T)$,
\begin{equation}
\frac{dF_T}{dC}<0,
\quad {\rm if}\quad
C<C_{\min}(T),
\qquad
{\rm and}
\qquad
\frac{dF_T}{dC}>0,
\quad {\rm if}\quad
C>C_{\min}(T).
\end{equation}

The position of the minimum is determined implicitly by
Eq.~\eqref{eq:dFT_dC_factorized},
\begin{equation}
\frac{
C_{\min}^2(T)\left[C_{\min}(T)+Q_\infty(T)+P_\infty(T)\right]
}{
2
\left[
\left(Q_\infty(T)+\frac{C_{\min}(T)}{2}\right)
\left(P_\infty(T)+\frac{C_{\min}(T)}{2}\right)
\right]^{3/2}
}
=1 .
\label{eq:Cmin_condition}
\end{equation}

To identify appropriate initial temperatures $T_{\rm h}$ and
$T_{\rm c}$ satisfying
${\cal D}_{\rm h}(0)>{\cal D}_{\rm c}(0)$, we do not attempt to
characterize all possible pairs, but restrict ourselves to a
sufficient class of preparations for which both initial covariance
matrices lie on the increasing branch of $F_T(C)$, namely
\begin{equation}
C_{\rm h}>C_{\rm c}>C_{\min}(T).
\label{eq:C_initial_class}
\end{equation}

It is then useful to introduce the temperature $T_{\rm init}(T)$
through
\begin{equation}
C\!\left(T_{\rm init}(T)\right)
=
\coth\left(
\frac{\omega_0}{2k_{\rm B}T_{\rm init}(T)}
\right)
=
C_{\min}(T),\quad\implies\quad T_{\rm init}(T)
=
\frac{\omega_0}
{2k_{\rm B}\operatorname{arccoth}[C_{\min}(T)]}.
\label{eq:Tinit_definition}
\end{equation}

It is also possible to show that, for any finite system--bath
coupling, the temperature $T_{\rm init}(T)$ defined above always
satisfies
\begin{equation}
T_{\rm init}(T)>T.
\label{eq:T_less_Tinit}
\end{equation}
Without entering into the details of the proof, the key ingredients
are the fluctuation--dissipation representation
\eqref{eq:QP_FDT}, the positivity of
${\rm Im}\,\tilde\chi(\omega)$, and the susceptibility sum rules
\begin{equation}
\frac{2}{\pi}
\int_0^\infty d\omega\,
\omega\,{\rm Im}\,\tilde\chi(\omega)=1,
\qquad
\frac{2}{\pi}
\int_0^\infty d\omega\,
\frac{{\rm Im}\,\tilde\chi(\omega)}{\omega}
=
\frac{1}{\omega_0^2}.
\label{eq:chi_sum_rules}
\end{equation}
These relations imply
\begin{equation}
Q_\infty(T)P_\infty(T)
>
\frac{1}{4}
\coth^2\left(\frac{\omega_0}{2k_{\rm B}T}\right)
=
\frac{C^2(T)}{4},
\label{eq:QP_inequality}
\end{equation}
where equality is recovered in the vanishing-coupling limit.
Using Eq.~\eqref{eq:QP_inequality} in
Eq.~\eqref{eq:dFT_dC_factorized}, one finds that the quantity in
curly brackets is still negative at $C=C(T)$. Since it is a strictly
increasing function of $C$ and vanishes at $C=C_{\min}(T)$, this
implies
\begin{equation}
C_{\min}(T)>C(T)\quad\implies\quad T_{\rm init}(T)>T,
\end{equation}
where in the last inequality we used the monotonicity of $C(T)$.
The condition \eqref{eq:C_initial_class} is therefore equivalent to
\begin{equation}
T<T_{\rm init}(T)<T_{\rm c}<T_{\rm h},
\label{eq:initial_temperature_class}
\end{equation}
which provides a sufficient condition for the required initial
ordering
$\mathcal D_{\rm h}(0)>\mathcal D_{\rm c}(0)$
within the restricted class considered here.
We stress that this condition is sufficient, but not necessary:
additional pairs satisfying the required initial ordering may exist
with $C_{\rm c}<C_{\min}(T)<C_{\rm h}$.

\section{Discussion of the $k_B T < \lambda_3/2\pi$ regime}
\label{sec:low_T_regime}
In this Section, we discuss the low-temperature regime $k_B T < \lambda_3/2\pi$, not reported in the main text, restricting for simplicity to the OD moderate-coupling and UD strong-coupling regimes.
In this temperature regime, the first Matsubara rate satisfies $A_1 = 2\pi k_B T < \lambda_3$, implying that the term $\Delta I_{ijn}^{(1)}(t)$ can no longer be neglected as a subleading term and must be retained. By keeping the leading asymptotic contributions, namely $i=3$ in $\Delta I_{ijn}^{(1)}(t)$ and $i=j=3$ in $\Delta I_{ijn}^{(2)}(t)$, the position and momentum fluctuations acquire an additional Matsubara-driven transient component,
\begin{align}
\langle \Delta Q(t)\rangle_{\mathrm{th}}^{(1)} &\simeq \sum_{n=0}^{\infty} Q_n e^{-(\lambda_3+A_n)t}, \\
\langle \Delta P(t)\rangle_{\mathrm{th}}^{(1)} &\simeq \sum_{n=0}^{\infty} P_n e^{-(\lambda_3+A_n)t},
\end{align}
where $P_n = (\lambda_3 A_n / \omega_0^2) Q_n$ and the coefficients $Q_n$ are given by
\begin{equation}
\begin{split}
    Q_n =& \frac{2\omega_0\chi_3 L_n}{m(A_n+\lambda_3)} \sum_{j=1}^{3} \frac{\chi_j}{A_n-\lambda_j}.\\
\end{split}
\label{eq:BQ}
\end{equation}
The sum in $Q_n$ is evaluated in sec. \ref{sec:sums}.

Consequently, the time-dependent part of the covariance matrix $\Delta\sigma_{\alpha}(t) = \sigma_{\mathrm{h},\alpha}(t) + \Delta\sigma_{\mathrm{th}}^{(1)}(t) + \Delta\sigma_{\mathrm{th}}^{(2)}(t)$ generalizes to
\begin{equation}
\Delta\sigma_{\alpha}(t) = e^{-2\lambda_3 t} \left[B_\alpha + B_{\mathrm{th}}\right] \mathbb{P} + \sum_{n=0}^{\infty} Q_n e^{-(\lambda_3+A_n) t} \,\mathbb{R}_n ,
\end{equation}
where the matrices $\mathbb{P}$ and $\mathbb{R}_n$ are defined as
\begin{equation}
\mathbb{P} = \begin{pmatrix} 
1 & -\lambda_3/\omega_0 \\ 
-\lambda_3/\omega_0 & (\lambda_3/\omega_0)^2 
\end{pmatrix}, 
\quad 
\mathbb{R}_n = \begin{pmatrix} 
1 & -\frac{\lambda_3+A_n}{2\omega_0} \\ 
-\frac{\lambda_3+A_n}{2\omega_0} & \frac{\lambda_3 A_n}{\omega_0^2} 
\end{pmatrix}.
\end{equation}

To determine the asymptotic behavior of the distance estimator $\mathcal{D}_{\alpha}(t)$, we examine the matrix $y_{\alpha}(t) = \sigma^{-1}(\infty)\Delta\sigma_{\alpha}(t)$. Since $A_1 < \lambda_3$, the two slowest decaying terms entering $y_{\alpha}(t)$ are proportional to $e^{-(\lambda_3+A_1)t}$ and $e^{-2\lambda_3 t}$. Truncating to these two leading contributions yields
\begin{equation}
y_\alpha(t) \approx e^{-(\lambda_3+A_1)t} Q_1 \mathbb{M}_1 + e^{-2\lambda_3 t} \left[B_\alpha+B_{\mathrm{th}}\right]\mathbb{M}_0 ,
\end{equation}
with $\mathbb{M}_0 = \sigma^{-1}(\infty)\mathbb{P}$ and $\mathbb{M}_1 = \sigma^{-1}(\infty)\mathbb{R}_1$.

Crucially, while the Matsubara contribution $e^{-(\lambda_3+A_1)t}$ represents the absolute slowest decay rate ($\lambda_3 + A_1 < 2\lambda_3$), the coefficient $Q_1$ depends solely on the bath parameters and is {independent} of the initial state temperature $T_\alpha$. 


Evaluating the Mpemba estimator $\Delta\mathcal{D}(t) = \mathcal{D}_{\mathrm{c}}(t) - \mathcal{D}_{\mathrm{h}}(t)$, the slowest term -- independent of the initial temperature -- cancels out completely.

The leading non-vanishing contribution to $\Delta\mathcal{D}(t)$ is therefore governed by the homogeneous scale $e^{-2\lambda_3 t}$, leading to
\begin{equation}
\Delta\mathcal{D}(t) \simeq \frac{\mathcal{C}_{1}}{D_{\infty}^{1/4}} \left(B_{\mathrm{c}} - B_{\mathrm{h}}\right) e^{-2\lambda_{3}t} ,
\label{eq:DeltaDLowT}
\end{equation}
where 
\begin{equation}
\mathcal{C}_1 = \frac{3}{4} \frac{\langle Q^2(\infty)\rangle^{-2} + \lambda_3(\lambda_3+A_1)\left(\omega_0^2 D_{\infty}\right)^{-1} + \lambda_3^3 A_1\left(\omega_0^2\langle P^2(\infty)\rangle\right)^{-2}}{\left\{ 3\left[ \langle Q^2(\infty)\rangle^{-2} + \lambda_3^2 A_1^2 \left(\omega_0^2\langle P^2(\infty)\rangle\right)^{-2} \right] + \left[ \lambda_3^2 + 4\lambda_3 A_1 + A_1^2 \right] \left(\omega_0^2 D_{\infty}\right)^{-1} \right\}^{1/2}} > 0
\end{equation}
is a positive constant depending on the stationary state.

Since $B_\alpha = \frac{1}{2} C_\alpha \chi_3^2 (\omega_0^2 + \lambda_3^2)$ is a monotonically increasing function of the initial temperature $T_\alpha$ via the initial covariance $C_\alpha = \coth(\beta_\alpha \omega_0 / 2)$, for a cold and hot initial state ($T_{\mathrm{c}} < T_{\mathrm{h}}$) we have $B_{\mathrm{c}} - B_{\mathrm{h}} < 0$. This directly implies
$
\Delta\mathcal{D}(t) < 0$.
We thus conclude that in the temperature regime $k_B T < \lambda_3/2\pi$, the direct Mpemba effect can't occur, as the distance trajectory of the hotter sample remains strictly bounded below that of the colder sample at long times.




\section{Compatibility between the initial ordering and the asymptotic inversion}
\label{app:compatibility}

We now combine the condition for the asymptotic inversion derived in
the main text with the requirement on the initial ordering discussed
in Sec.~\ref{app:initial_distance}. In the overdamped and strong-coupling
regimes, the asymptotic inversion requires
\begin{equation}
B_{\rm th}(T)
<
-\frac{\chi_3^2(\omega_0^2+\lambda_3^2)}{4}
\left[C_{\rm c}+C_{\rm h}\right].
\label{eq:asymptotic_condition_compatibility}
\end{equation}
As shown in the main text, there exist temperatures
$T_{\rm h}>T_{\rm c}>T$ satisfying
Eq.~\eqref{eq:asymptotic_condition_compatibility}
if and only if
\begin{equation}
T>T^*.
\label{eq:Tstar_condition_compatibility}
\end{equation}

On the other hand, within the sufficient class of initial
preparations selected in Sec.~\ref{app:initial_distance}, the required
initial ordering
${\cal D}_{\rm h}(0)>{\cal D}_{\rm c}(0)$
is guaranteed by
\begin{equation}
T<T_{\rm init}(T)<T_{\rm c}<T_{\rm h}.
\label{eq:initial_condition_compatibility}
\end{equation}

A direct Mpemba effect within this class therefore requires the
simultaneous fulfillment of
Eqs.~\eqref{eq:asymptotic_condition_compatibility}
and \eqref{eq:initial_condition_compatibility}.
We therefore examine whether the two sets of conditions have a
nonempty intersection.

For fixed $T$ and $T_{\rm c}$, the asymptotic condition (\ref{eq:asymptotic_condition_compatibility}) becomes
progressively more restrictive as $T_{\rm h}$ increases, since
$C_{\rm h}$ is a monotonically increasing function of $T_{\rm h}$.
Hence, for a given $T_{\rm c}$, the most favorable limit is
$T_{\rm h}\rightarrow T_{\rm c}^{+}$.
This motivates the definition of a maximal cold-preparation
temperature $T_{\rm c}^{\max}(T)$ through
\begin{equation}
C\!\left[T_{\rm c}^{\max}(T)\right]
=
-\frac{2B_{\rm th}(T)}
{\chi_3^2(\omega_0^2+\lambda_3^2)}.
\label{eq:Tcmax_definition}
\end{equation}

Then, for
\begin{equation}
T_{\rm c}<T_{\rm c}^{\max}(T),
\end{equation}
there exists at least one $T_{\rm h}>T_{\rm c}$ satisfying the
asymptotic inversion condition.
Combining this with the initial-ordering requirement
$T_{\rm c}>T_{\rm init}(T)$, the two conditions can be simultaneously
fulfilled whenever
\begin{equation}
T_{\rm init}(T)<T_{\rm c}<T_{\rm c}^{\max}(T).
\label{eq:Tc_allowed_interval}
\end{equation}

Our aim is now to establish that the selected class of initial
preparations compatible with the asymptotic inversion is not empty.
According to Eq.~\eqref{eq:Tc_allowed_interval}, this amounts to
showing that the interval
$
T_{\rm init}(T)<T_{\rm c}<T_{\rm c}^{\max}(T)$
is nonempty. It is therefore convenient to define
\begin{equation}
G(T)=T_{\rm c}^{\max}(T)-T_{\rm init}(T).
\label{eq:G_definition}
\end{equation}
The required condition is simply $G(T)>0$.

To establish the existence of such a region, it is sufficient to
consider the high-temperature limit. In this regime the stationary
variances approach their classical values,
$
Q_\infty=P_\infty=
\frac{k_{\rm B}T}{\omega_0}+O(T^{-1})$.
Inserting these expressions into the equation (\ref{eq:Cmin_condition}) gives
$
C_{\min}(T)\simeq 2k_{\rm B}T/\omega_0$.
Since $C(T_r)\simeq 2k_{\rm B}T_r/\omega_0$ at high temperature, the
definition of $T_{\rm init}(T)$ then yields
\begin{equation}
T_{\rm init}(T)\simeq T.
\label{eq:Tinit_highT}
\end{equation}

We next consider $T_{\rm c}^{\max}(T)$. In the same high-temperature
limit, the thermal contribution entering the asymptotic condition
behaves as
\begin{equation}
B_{\rm th}(T)
\simeq
-\chi_3^2(\omega_0^2+\lambda_3^2)
\frac{\omega_0 k_{\rm B}T}{\lambda_3^2},
\label{eq:Bth_highT}
\end{equation}
where the characteristic equation for the pole $\lambda_3$ has been
used. From Eq.~\eqref{eq:Tcmax_definition} it follows that
$
C\!\left(T_{\rm c}^{\max}(T)\right)
\simeq 2\omega_0 k_{\rm B}T/\lambda_3^2$.
Using again the high-temperature form of $C(T_r)$, we obtain
\begin{equation}
T_{\rm c}^{\max}(T)
\simeq
\frac{\omega_0^2}{\lambda_3^2}T.
\label{eq:Tcmax_highT}
\end{equation}

Equations~\eqref{eq:Tinit_highT} and \eqref{eq:Tcmax_highT} therefore
give at high temperatures 
\begin{equation}
G(T)
\simeq
\left(\frac{\omega_0^2}{\lambda_3^2}-1\right)T.
\label{eq:G_highT}
\end{equation}
In the overdamped and strong-coupling regimes considered here,
$\lambda_3<\omega_0$, so that at high $T$ it is $G(T)>0$. 
Consequently, there exists a finite temperature $\widetilde T$ such
that
\begin{equation}
G(T)>0,
\qquad {\rm for}\quad T>\widetilde T .
\end{equation}
Hence, for sufficiently large bath temperatures, there exists a
nonempty interval
\begin{equation}
T_{\rm init}(T)<T_{\rm c}<T_{\rm c}^{\max}(T),
\end{equation}
whose lower bound guarantees the required initial ordering, while its
upper bound ensures that the asymptotic inversion condition can still
be satisfied. For any $T_{\rm c}$ chosen in this interval, one can
then choose a sufficiently close $T_{\rm h}>T_{\rm c}$ such that both
requirements are simultaneously fulfilled. The selected class of
preparations realizing the direct Mpemba effect is therefore
nonempty.

\section{Initial time slip and Mpemba effect} 
We now detail the proof that the emergence of the Mpemba effect is due to the initial time slip term in the quantum Langevin equation.

All the results presented in the main text have been obtained from the exact Langevin equation, whose homogeneous part is
\begin{equation}
\ddot Q(t)+\omega_0^2 Q(t)
+\int_0^t d\tau\,\gamma(t-\tau)\dot Q(\tau)
+\gamma(t)Q(0)=0.
\label{eq:Langevin_slip}
\end{equation}
The last contribution, proportional to the initial coordinate \(Q(0)\), is
the {initial-slip} (or {initial time-slip}) term. Such a
term naturally arises in the exact Caldeira--Leggett model when the total
initial state is assumed to be factorized into an uncorrelated system state
and a thermal bath state. Physically, it describes the rapid
build-up of system--bath correlations immediately after the interaction is
switched on.

The initial-slip contribution affects the dynamics only through the dependence
on the initial conditions and therefore modifies exclusively the homogeneous
part of the solution. By contrast, the thermal contribution is entirely
determined by the bath-noise correlations through the response function
\(\chi(t)\), which is unaffected by the presence or absence of the slip term.

For this reason, assessing the role of the initial-slip term requires
reconsidering only the homogeneous dynamics. We therefore formally remove the
contribution \(\gamma(t)Q(0)\) from Eq.~\eqref{eq:Langevin_slip}. The
homogeneous Langevin equation then reduces to
\begin{equation}
\ddot Q_{\rm h}(t)+\omega_0^2 Q_{\rm h}(t)
+\int_0^t d\tau\,\gamma(t-\tau)\dot Q_{\rm h}(\tau)=0.
\label{eq:Langevin_no_slip}
\end{equation}

In the following, we compare the dynamics generated by
Eq.~\eqref{eq:Langevin_no_slip} with that obtained from the full equation
\eqref{eq:Langevin_slip}, in order to determine whether the quantum Mpemba
effect is robust against the removal of the initial-slip contribution.
Taking the Laplace transform of Eq.~\eqref{eq:Langevin_no_slip},
\begin{equation}
\widehat Q_{\rm h}(\lambda)
=\int_0^\infty dt\,e^{-\lambda t}Q_{\rm h}(t),
\end{equation}
and using the standard relations for the Laplace transform of time
derivatives, we obtain
\begin{equation}
\left[
\lambda^2+\lambda\widehat\gamma(\lambda)+\omega_0^2
\right]
\widehat Q_{\rm h}(\lambda)
=
\left[
\lambda+\widehat\gamma(\lambda)
\right]Q(0)
+\omega_0 P(0).
\label{eq:Q_laplace}
\end{equation}
It is convenient to express this result in terms of the response function
introduced in the main text, whose Laplace transform is
\begin{equation}
\widehat\chi(\lambda)
=
\frac{1}
{\lambda^2+\lambda\widehat\gamma(\lambda)+\omega_0^2}.
\label{eq:chi_lambda}
\end{equation}
The corresponding time-domain response function satisfies the initial
conditions
\begin{equation}
\chi(0)=0,
\qquad
\dot\chi(0)=1.
\end{equation}
Substituting Eq.~\eqref{eq:chi_lambda} into
Eq.~\eqref{eq:Q_laplace}, the homogeneous solution can be written as
\begin{equation}
\widehat Q_{\rm h}(\lambda)
=
\widehat G(\lambda)\,Q(0)
+\omega_0\,\widehat\chi(\lambda)\,P(0),
\label{eq:Qh_laplace_compact}
\end{equation}
where we have introduced
$
\widehat G(\lambda)
=
\left[\lambda+\widehat\gamma(\lambda)\right]
\widehat\chi(\lambda).
$
Using Eq.~\eqref{eq:chi_lambda}, this expression can be rewritten in the
equivalent form
\begin{equation}
\widehat G(\lambda)
=
\frac{1-\omega_0^2\widehat\chi(\lambda)}{\lambda}.
\label{eq:G_lambda_identity}
\end{equation}
The inverse Laplace transform of
Eq.~\eqref{eq:Qh_laplace_compact} yields
\begin{equation}
Q_{\rm h}^{\rm ns}(t)
=
G(t)\,Q(0)
+\omega_0\,\chi(t)\,P(0),
\label{eq:Qh_no_slip}
\end{equation}
where the superscript ``ns'' indicates the absence of the initial-slip term.
The function \(G(t)\) is defined as
\begin{equation}
G(t)
=
1-\omega_0^2\int_0^t d\tau\,\chi(\tau).
\label{eq:G_integral}
\end{equation}
It is useful to rewrite Eq.~\eqref{eq:G_integral} as
\begin{align}
G(t)
&=
1-\omega_0^2\int_0^\infty d\tau\,\chi(\tau)
+\omega_0^2\int_t^\infty d\tau\,\chi(\tau) =
\omega_0^2\int_t^\infty d\tau\,\chi(\tau),
\label{eq:G_tail}
\end{align}
where in the second line we have used the identity
\begin{equation}
\int_0^\infty d\tau\,\chi(\tau)
=
\widehat\chi(0)
=
\frac{1}{\omega_0^2}.
\label{eq:chi_zero_identity}
\end{equation}
Using the exponential decomposition of the response function,
\begin{equation}
\chi(t)
=
\sum_{i=1}^3 \chi_i e^{-\lambda_i t},
\label{eq:chi_mode_decomposition}
\end{equation}
Eq.~\eqref{eq:G_tail} immediately gives
\begin{equation}
G(t)
=
\omega_0^2
\sum_{i=1}^3
\frac{\chi_i}{\lambda_i}
e^{-\lambda_i t}.
\label{eq:G_three_modes}
\end{equation}
The evaluation of the momentum operator \(P_{\rm h}^{\rm ns}(t)\) follows directly
from the relation
\begin{equation}
P_{\rm h}^{\rm ns}(t)=\frac{\dot Q_{\rm h}^{\rm ns}(t)}{\omega_0}.
\end{equation}
Using Eq.~\eqref{eq:Qh_no_slip}, we obtain
\begin{equation}
P_{\rm h}^{\rm ns}(t)
=
-\omega_0\,\chi(t)\,Q(0)
+\dot\chi(t)\,P(0).
\label{eq:Ph_no_slip}
\end{equation}
For comparison, we recall the corresponding homogeneous solutions in the
presence of the initial-slip term, namely
\begin{equation}
Q_{\rm h}(t)
=
\dot\chi(t)\,Q(0)
+\omega_0\,\chi(t)\,P(0),
\label{eq:Qh_slip}
\end{equation}
and
\begin{equation}
P_{\rm h}(t)
=
\frac{\ddot\chi(t)}{\omega_0}\,Q(0)
+\dot\chi(t)\,P(0).
\label{eq:Ph_slip}
\end{equation}
The only difference between the two homogeneous solutions is the function
multiplying the initial position \(Q(0)\): in the absence of the initial-slip
term, \(\dot\chi(t)\) is replaced by \(G(t)\).

We assume that the oscillator is initially prepared in a thermal state at
temperature \(T_\alpha\). The initial covariance matrix is therefore diagonal,
with elements
\begin{equation}
\langle Q^2(0)\rangle_\alpha
=
\langle P^2(0)\rangle_\alpha
=
\frac{C_\alpha}{2},
\qquad
\left\langle
Q(0)P(0)+P(0)Q(0)
\right\rangle_\alpha
=
0.
\label{eq:initial_covariance_ns}
\end{equation}
The homogeneous position and momentum variances then read
\begin{align}
\left\langle Q_{\rm h}^{\rm ns}(t)^2 \right\rangle_\alpha
&=
\frac{C_\alpha}{2}
\left[
G^2(t)+\omega_0^2\chi^2(t)
\right],
\label{eq:Q_variance_ns}
\\[6pt]
\left\langle P_{\rm h}^{\rm ns}(t)^2 \right\rangle_\alpha
&=
\frac{C_\alpha}{2}
\left[
\dot\chi^2(t)+\omega_0^2\chi^2(t)
\right].
\label{eq:P_variance_ns}
\end{align}

\subsection{From the moderate to the strong coupling regime without the initial-slip term}
To further investigate the effect of removing the initial-slip contribution, we
consider the moderate-to-strong coupling regime, where the slowest decay mode
$\lambda_3$ is real. In the long-time limit, retaining only the contribution
associated with this slowest mode, we obtain
\begin{equation}
G(t)
\simeq
\frac{\omega_0^2}{\lambda_3}
\chi_3 e^{-\lambda_3 t}.
\label{eq:G_lambda3}
\end{equation}
Consequently, Eq.~\eqref{eq:Qh_no_slip} becomes
\begin{equation}
Q_{\rm h}^{\rm ns}(t)
\simeq
\omega_0\chi_3 e^{-\lambda_3 t}
\left[
\frac{\omega_0}{\lambda_3}Q(0)
+
P(0)
\right].
\label{eq:Qh_asymptotic}
\end{equation}
Similarly, using Eq.~\eqref{eq:Ph_no_slip} and the asymptotic form of
$\dot{\chi}(t)$, we obtain
\begin{equation}
P_{\rm h}^{\rm ns}(t)
\simeq
-\lambda_3\chi_3 e^{-\lambda_3 t}
\left[
\frac{\omega_0}{\lambda_3}Q(0)
+
P(0)
\right].
\label{eq:Ph_asymptotic}
\end{equation}
It follows that the homogeneous position variance is given by
\begin{equation}
\left\langle
Q_{\rm h}^{\rm ns}(t)^2
\right\rangle_\alpha
=
\frac{C_\alpha}{2}
\chi_3^2
\left(\frac{\omega_0}{\lambda_3}\right)^2
\left(\omega_0^2+\lambda_3^2\right)
e^{-2\lambda_3 t}.
\label{eq:QQ_h}
\end{equation}
Similarly, the momentum variance is given by
\begin{equation}
\left\langle
P_{\rm h}^{\rm ns}(t)^2
\right\rangle_\alpha
=
\frac{C_\alpha}{2}
\chi_3^2
(\omega_0^2+\lambda_3^2)
e^{-2\lambda_3 t}.
\label{eq:PP_h}
\end{equation}
Finally, the mixed covariance term reads
\begin{equation}
\frac12
\left\langle
\{Q_{\rm h}^{\rm ns}(t),P_{\rm h}^{\rm ns}(t)\}
\right\rangle_\alpha
=
-\frac{C_\alpha}{2}
\chi_3^2
\frac{\omega_0}{\lambda_3}
(\omega_0^2+\lambda_3^2)
e^{-2\lambda_3 t}.
\label{eq:QP_h}
\end{equation}
Collecting the above expressions, the homogeneous contribution to the
covariance matrix in the absence of the initial-slip term becomes
\begin{equation}
\sigma_{\rm h,\alpha}^{\rm ns}(t)
=
\left(\frac{\omega_0}{\lambda_3}\right)^2
\frac{C_\alpha}{2}
\chi_3^2
(\omega_0^2+\lambda_3^2)
e^{-2\lambda_3 t}
\left(
\begin{array}{cc}
1 &
-\lambda_3/\omega_0
\\
-\lambda_3/\omega_0 &
\lambda_3^2/\omega_0^2
\end{array}
\right).
\label{eq:sigmah_ns}
\end{equation}

This expression differs from the corresponding homogeneous covariance matrix
obtained in the presence of the initial-slip term only by the overall factor
\((\omega_0/\lambda_3)^2\) (see main text).
Therefore, in the long-time regime dominated by the slowest decay mode, the
two covariance matrices are related by
\begin{equation}
\sigma_{\mathrm{h},\alpha}^{\rm ns}(t)
=
\left(\frac{\omega_0}{\lambda_3}\right)^2
\sigma_{\mathrm{h},\alpha}(t).
\label{eq:sigma_relation}
\end{equation}
Since the homogeneous covariance matrices with and without the initial-slip
term are proportional to each other, the analysis developed in the main text
for determining the occurrence of the Mpemba effect remains unchanged. In
particular, the necessary and sufficient condition derived in
the main text retains the same form, with the replacement of the
coefficients \(B_\alpha\) by their counterparts \(B_\alpha^{\rm ns}\). The
condition therefore becomes
\begin{equation}
B_{\mathrm{th}}(T)
<
-\frac{
B^{\rm ns}_{\mathrm{c}}(T_{\mathrm c})
+
B^{\rm ns}_{\mathrm{h}}(T_{\mathrm h})
}{2}.
\label{condizione1}
\end{equation}
Here, the thermal contribution \(B_{\mathrm{th}}(T)\) is unchanged (see End Matter), whereas the homogeneous contribution becomes
\begin{equation}
B^{\rm ns}_{\alpha}(T_\alpha)
=
\chi_3^2
\left(\frac{\omega_0}{\lambda_3}\right)^2
\frac{\omega_0^2+\lambda_3^2}{2}
\coth\!\left(
\frac{\omega_0}{2k_{\mathrm B}T_\alpha}
\right).
\label{eq:MpembaCondition}
\end{equation}

As shown in the End Matter, the necessary condition for
satisfying Eq.~\eqref{condizione1} can be expressed as
${\cal A}>{\cal B}^{\rm ns},$
where the two quantities are defined as
\begin{equation}
{\cal A}
=
-\lim_{T\rightarrow\infty}
\frac{B_{\rm th}(T)}{k_{\rm B}T},
\qquad
{\cal B}^{\rm ns}
=
\lim_{T\rightarrow\infty}
\frac{
B_{\rm c}^{\rm ns}(T)
+
B_{\rm h}^{\rm ns}(T)
}
{2k_{\rm B}T}.
\end{equation}
Substituting the explicit expressions for the different contributions gives
\begin{equation}
{\cal A}
=
\frac{
\chi_3^2
\gamma_0
\omega_D
\omega_0
}
{\lambda_3(\omega_D-\lambda_3)},
\qquad
{\cal B}^{\rm ns}
=
\chi_3^2
\left(\frac{\omega_0}{\lambda_3}\right)^2
\frac{\omega_0^2+\lambda_3^2}{\omega_0}.
\end{equation}
The ratio between the two quantities is therefore
\begin{equation}
\frac{{\cal A}}{{\cal B}^{\rm ns}}
=
\frac{
\lambda_3\gamma_0\omega_D
}
{
(\omega_D-\lambda_3)
(\omega_0^2+\lambda_3^2)
}.
\end{equation}

Using the fact that \(\lambda_3\) is a root of the cubic equation
\(P(\lambda)=0\), we have
$
\lambda_3\gamma_0\omega_D
=
(\omega_D-\lambda_3)
(\omega_0^2+\lambda_3^2).
$
Consequently,
$
{{\cal A}}/{{\cal B}^{\rm ns}}=1.
$
Therefore, the necessary condition for the occurrence of the Mpemba effect,
which requires the strict inequality \({\cal A}>{\cal B}^{\rm ns}\), is never
satisfied. This demonstrates that the Mpemba effect cannot emerge in the
absence of the initial-slip term.

\subsection{Underdamped weak-coupling regime without the initial-slip term}
We now analyze the Mpemba effect in the underdamped
weak-coupling regime, which is particularly relevant for comparison with
approximate descriptions based on quantum master equations and Lindblad
dynamics. As discussed in the main text, for
\(\gamma_0<2\omega_0\) the two slowest decay modes form a complex-conjugate
pair,
\begin{equation}
\lambda_{2,3}=\lambda\pm i\eta ,
\end{equation}
with corresponding residues satisfying
\(\chi_2=\chi_3^*\). In this regime, the long-time behavior of the function
\(G(t)\) is dominated by these two modes and reads
\begin{equation}
G(t)
\simeq
\omega_0^2 e^{-\lambda t}
\left[
\frac{\chi_3}{\lambda_3}e^{-i\eta t}
+\mathrm{c.c.}
\right].
\label{1eq:G_lambda3}
\end{equation}

Using Eqs.~\eqref{eq:Qh_no_slip} and~\eqref{eq:Ph_no_slip}, the homogeneous
contribution to the covariance matrix can be written as
\begin{align}
\left\langle Q_{\rm h}^{\rm ns}(t)^2\right\rangle_\alpha
&=
\frac{C_\alpha}{2}\omega_0^2e^{-2\lambda t}
\Bigg[
\frac{|\chi_3|^2}{|\lambda_3|^2}
(\omega_0^2+|\lambda_3|^2)
+
\frac{\chi_3^2}{\lambda_3^2}
(\omega_0^2+\lambda_3^2)e^{-2i\eta t}
+\mathrm{c.c.}
\Bigg],
\nonumber\\
\left\langle P_{\rm h}^{\rm ns}(t)^2\right\rangle_\alpha
&=
\frac{C_\alpha}{2}e^{-2\lambda t}
\Bigg[
|\chi_3|^2(\omega_0^2+|\lambda_3|^2)
+
\chi_3^2(\omega_0^2+\lambda_3^2)e^{-2i\eta t}
+\mathrm{c.c.}
\Bigg],
\nonumber\\
\frac12
\left\langle
\{Q_{\rm h}^{\rm ns}(t),P_{\rm h}^{\rm ns}(t)\}
\right\rangle_\alpha
&=
-\frac{C_\alpha}{2}\omega_0e^{-2\lambda t}
\Bigg[
\frac{\lambda|\chi_3|^2}{|\lambda_3|^2}
(\omega_0^2+|\lambda_3|^2)
+
\frac{\chi_3^2}{\lambda_3}
(\omega_0^2+\lambda_3^2)e^{-2i\eta t}
+\mathrm{c.c.}
\Bigg].
\label{1eq:PP_h}
\end{align}

We now rewrite the thermal contribution in the weak-coupling regime. It takes the form
\begin{align}
\langle \Delta Q(t)\rangle_{\mathrm{th}}
&=
-\frac{\omega_0}{m}e^{-2\lambda t}
\Bigg[
\frac{|\chi_3|^2}{\lambda}S_1(\lambda_3)
+
\frac{\chi_3^2}{\lambda_3}
S_1(\lambda_3)e^{-2i\eta t}
+\mathrm{c.c.}
\Bigg],
\nonumber\\
\langle \Delta P(t)\rangle_{\mathrm{th}}
&=
-\frac{1}{m\omega_0}e^{-2\lambda t}
\Bigg[
\frac{|\chi_3|^2|\lambda_3|^2}{\lambda}
S_1(\lambda_3)
+
\chi_3^2\lambda_3
S_1(\lambda_3)e^{-2i\eta t}
+\mathrm{c.c.}
\Bigg],
\nonumber\\
\frac{d}{2\omega_0dt}
\langle \Delta Q(t)\rangle_{\mathrm{th}}
&=
\frac{1}{m}e^{-2\lambda t}
\Big[
|\chi_3|^2S_1(\lambda_3)
+
\chi_3^2S_1(\lambda_3)e^{-2i\eta t}
+\mathrm{c.c.}
\Big].
\label{weakthermal}
\end{align}
where, according to Eq.~\eqref{S1},
\begin{equation}
S_1(\lambda_3)
=
\frac{m\gamma_0\omega_D}{\omega_D-\lambda_3}
\left\{
k_BT
+
\frac{\lambda_3\omega_D}
{\pi(\omega_D+\lambda_3)}
\left[
\psi\left(1-\frac{\beta\lambda_3}{2\pi}\right)
-
\psi\left(1+\frac{\beta\omega_D}{2\pi}\right)
\right]
\right\}.
\end{equation}
Since \(\lambda_3\) is a root of the characteristic cubic equation, it
satisfies
\begin{equation}
\lambda_3\gamma_0\omega_D
=
(\omega_D-\lambda_3)(\omega_0^2+\lambda_3^2).
\end{equation}
Using this identity, \(S_1(\lambda_3)\) can be rewritten as
\begin{equation}
S_1(\lambda_3)
=
\frac{m(\omega_0^2+\lambda_3^2)}{\lambda_3}
\left\{
k_BT
+
\frac{\lambda_3\omega_D}
{\pi(\omega_D+\lambda_3)}
\left[
\psi\left(1-\frac{\beta\lambda_3}{2\pi}\right)
-
\psi\left(1+\frac{\beta\omega_D}{2\pi}\right)
\right]
\right\}
\equiv
\frac{m(\omega_0^2+\lambda_3^2)}{\lambda_3}
F(\lambda_3).
\end{equation}
Combining the homogeneous and thermal contributions, we obtain
\begin{align}
\left\langle Q_{\rm h}^{\rm ns}(t)^2\right\rangle_\alpha
+\langle\Delta Q(t)\rangle_{\mathrm{th}}
&=
\omega_0e^{-2\lambda t}
\Bigg\{
\frac{\omega_0^2+\lambda_3^2}{\lambda_3}
\left[
\frac{\omega_0C_\alpha}{2}-F(\lambda_3)
\right]
\left[
\frac{|\chi_3|^2}{\lambda}
+
\frac{\chi_3^2}{\lambda_3}e^{-2i\eta t}
\right]
+\mathrm{c.c.}
\Bigg\},
\nonumber\\
\left\langle P_{\rm h}^{\rm ns}(t)^2\right\rangle_\alpha
+\langle\Delta P(t)\rangle_{\mathrm{th}}
&=
\frac{e^{-2\lambda t}}{\omega_0}
\Bigg\{
\frac{\omega_0^2+\lambda_3^2}{\lambda_3}
\left[
\frac{\omega_0C_\alpha}{2}-F(\lambda_3)
\right]
\left[
\frac{|\chi_3|^2|\lambda_3|^2}{\lambda}
+
\chi_3^2\lambda_3e^{-2i\eta t}
\right]
+\mathrm{c.c.}
\Bigg\},
\nonumber\\
\frac{d}{2\omega_0dt}
\left[
\left\langle Q_{\rm h}^{\rm ns}(t)^2\right\rangle_\alpha
+\langle\Delta Q(t)\rangle_{\mathrm{th}}
\right]
&=
-e^{-2\lambda t}
\Bigg\{
\frac{\omega_0^2+\lambda_3^2}{\lambda_3}
\left[
\frac{\omega_0C_\alpha}{2}-F(\lambda_3)
\right]
\left[
|\chi_3|^2
+
\chi_3^2e^{-2i\eta t}
\right]
+\mathrm{c.c.}
\Bigg\}.
\label{weakthermal1}
\end{align}
These exact expressions provide the starting point for analyzing the difference
between the Frobenius distances. A fully analytical derivation of the condition
for the occurrence of the Mpemba effect is, however, possible only in the
Markovian limit, which will be discussed in the following subsection. Away from
this limit, in particular at lower temperatures, the analysis necessarily
requires numerical evaluation.

The numerical results presented below are obtained for temperatures
sufficiently larger than the characteristic scale
\(\lambda/(2\pi k_{\mathrm B})\), thus excluding the narrow low-temperature
region around the crossover condition
\(2\pi k_{\mathrm B}T=\lambda\). In the temperature range considered here, the
asymptotic dynamics remains well separated from this crossover region,
allowing us to clearly identify the effect of neglecting the initial-slip term
on the emergence of the Mpemba effect.
\begin{figure}[ht]
        \includegraphics[width=0.48\textwidth]{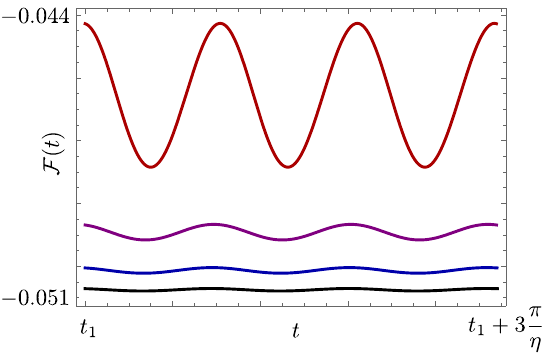}
        \caption{Plot of $\mathcal{F}(t)$ in the weak coupling regime without initial--slip contribution as a function of time $t$ (in units of $\omega_0^{-1}$) for $\omega_{\mathrm D}=10\omega_0$, $k_{\mathrm B}T=0.5\omega_0$, $k_{\mathrm{B}}T_{\mathrm c}=0.6\omega_0$, and $k_{\mathrm{B}}\Delta T=k_{\mathrm{B}}(T_{\mathrm h}-T_{\mathrm c})=0.05\omega_0$, for different values of the interaction strength: (black) $\gamma_0=0.04\omega_0$; (blue) $\gamma_0=0.06\omega_0$; (purple) $\gamma_0=0.1\omega_0$; (red) $\gamma_0=0.3\omega_0$.}
        \label{fig:FigDWeak_NoSlip}
\end{figure}

Figure~\ref{fig:FigDWeak_NoSlip} shows the behavior of
\(\mathcal{F}(t)=\Delta\mathcal{D}(t)e^{2\lambda t}\) for different values of
the coupling strength \(\gamma_0\). In all cases,
\(\mathcal{F}(t)\) remains negative, implying
\(\Delta\mathcal{D}(t)<0\), and therefore no Mpemba effect is observed. Extensive numerical analysis in the same temperature regime confirms that this
behavior is generic: in the absence of the initial-slip contribution, the Mpemba effect does not occur.

\subsection{Markovian limit}
\label{sec:markovian_limit}
We now consider the Markovian limit of the exact weak-coupling dynamics
derived in the previous subsection in the absence of the initial-slip
contribution. In the underdamped regime, this limit is obtained by assuming
the hierarchy of energy scales
$
k_{\rm B}T\gg \omega_D\gg \gamma_0,\omega_0 .
$
Under this condition, the bath correlation time becomes much shorter than the
characteristic relaxation times of the system. In this regime, the function
\(F(\lambda_3)\) can be simplified by using the high-temperature expansion of
the digamma functions, yielding
\begin{equation}
F(\lambda_3)\simeq k_{\rm B}T .
\label{eq:F_markovian}
\end{equation}

Furthermore, in the same limit
\(\omega_D\gg \gamma_0,\omega_0\), the two slowest decay modes reduce to the
standard underdamped Markovian poles,
\begin{equation}
\lambda=\frac{\gamma_0}{2},
\qquad
\eta^2=\omega_0^2-\lambda^2,
\qquad
\chi_3=\frac{i}{2\eta}.
\label{eq:markovian_roots}
\end{equation}
Substituting Eqs.~\eqref{eq:F_markovian} and
\eqref{eq:markovian_roots} into the exact covariance-matrix elements obtained
above, the asymptotic covariance matrix takes a considerably simpler form. In
particular, the temperature-dependent prefactor becomes
\begin{equation}
\frac{\omega_0}{2}C_\alpha-F(\lambda_3)
\;\longrightarrow\;
k_{\rm B}(T_\alpha-T),
\label{eq:Kalpha_markovian}
\end{equation}
which is real in this limit. Therefore, the covariance matrix can be written
as
\begin{equation}
\sigma^{\rm ns}_{\alpha}(t)
=
\sigma(\infty)+\Delta\sigma^{\rm ns}_{\alpha}(t)
=
\frac{k_{\rm B}T}{\omega_0}
+
\frac{k_{\rm B}(T_\alpha-T)}{\omega_0}
e^{-2\lambda t}
\mathbb{M}^{\rm ns}(t).
\label{sigmamarkov}
\end{equation}
Here, the first term represents the classical stationary covariance, while the
second term describes the asymptotic relaxation towards equilibrium. The
dimensionless matrix \(\mathbb{M}^{\rm ns}(t)\) is given by
\begin{equation}
\mathbb{M}^{\rm ns}(t)
=
\frac{1}{\eta^2}
\left(
\begin{array}{cc}
\omega_0^2-\lambda^2\cos(2\eta t)
+\lambda\eta\sin(2\eta t)
&
-\lambda\omega_0\left[1-\cos(2\eta t)\right]
\\[2mm]
-\lambda\omega_0\left[1-\cos(2\eta t)\right]
&
\omega_0^2-\lambda^2\cos(2\eta t)
-\lambda\eta\sin(2\eta t)
\end{array}
\right).
\label{eq:Mmarkov}
\end{equation}

This expression follows directly by substituting
Eqs.~\eqref{eq:F_markovian} and~\eqref{eq:markovian_roots} into the exact
covariance-matrix elements reported in Eq.~\eqref{weakthermal1}. The key
simplification relies on the identity
\begin{equation}
\frac{\omega_0^2+\lambda_3^2}{\lambda_3}
=
2\lambda ,
\label{eq:key_identity_markov}
\end{equation}
which follows from the Markovian relations
\(\lambda^2+\eta^2=\omega_0^2\) and
\(\lambda_3=\lambda+i\eta\).

Notice that, even without specifying the explicit form of
\(\mathbb{M}^{\rm ns}(t)\), the structure of the asymptotic
distance already reveals an important property. Indeed, evaluating the
distance through
\begin{equation}
\mathcal{D}_{\alpha}(t)
=
\frac{1}{4D_{\infty}^{1/4}}
\sqrt{
\tr^2\{y_{\alpha}(t)\}
+
2\,\tr\{y_{\alpha}^2(t)\}
},
\label{Dtracciam}
\end{equation}
one immediately finds that the dependence on the initial temperature
factorizes as
\begin{equation}
\mathcal{D}_{\alpha}(t)
=
\frac{k_{\rm B}(T_{\alpha}-T)e^{-2\lambda t}}
{4\omega_0D_{\infty}^{1/4}}
\sqrt{
\tr^2\{\sigma(\infty)^{-1}
\mathbb{M}^{\rm ns}(t)\}
+
2\,\tr\{
[\sigma(\infty)^{-1}
\mathbb{M}^{\rm ns}(t)]^2
\}
}.
\label{Dtracciam1}
\end{equation}

Therefore, the entire time dependence is independent of the initial
temperature, and the difference between the asymptotic distances associated
with the cold and hot initial states can always be written as
\begin{equation}
D_{\rm c}(t)-D_{\rm h}(t)
=
(T_{\rm c}-T_{\rm h})\,\mathcal{G}(t),
\end{equation}
where \(\mathcal{G}(t)>0\). Since \(T_{\rm h}>T_{\rm c}\), it follows that
$D_{\rm c}(t)<D_{\rm h}(t)
$
at all times. Hence, the two relaxation curves never cross and the Mpemba
effect is absent in the Markovian limit of the exact dynamics without the
initial-slip contribution.

We emphasize that this conclusion follows directly from the exact
weak-coupling solution after removing the initial-slip term and does not
involve any additional approximation based on a quantum master equation or
Lindblad description.

\subsubsection{Comparison with the Born--Markov master equation}
We now compare the above result with the one obtained from the standard
Born--Markov master equation derived in Ref.~\cite{Petruccione}, within the
same hierarchy of energy scales,
$
k_{\rm B}T\gg \omega_D\gg \omega_0,\gamma_0,
\,\,
\omega_0>\gamma_0/2 .
$

We show that the two approaches lead to the same asymptotic behavior. In our
notation, the reduced density operator obtained within the Born--Markov
approximation satisfies~\cite{Petruccione}
\begin{equation}
\frac{d}{dt}\rho_{\rm osc}(t)
=
-\frac{i}{\hbar}
[H_{\rm osc},\rho_{\rm osc}(t)]
-\frac{i\gamma_0}{2\hbar}
[x,\{p,\rho_{\rm osc}(t)\}]
-\frac{m\gamma_0 k_{\rm B}T}{\hbar^2}
[x,[x,\rho_{\rm osc}(t)]] .
\label{eq:petruccione_master_equation}
\end{equation}

From this equation, one can directly derive the equations of motion for the
second moments. Introducing the notation
\begin{equation}
X(t)=\langle Q^2(t)\rangle_{\rm ME},
\qquad
Y(t)=\langle P^2(t)\rangle_{\rm ME},
\end{equation}
the corresponding equations become
\begin{align}
\ddot X(t)
&=
2\omega_0^2Y(t)
-2\omega_0^2X(t)
-\gamma_0\dot X(t),
\nonumber\\
\dot Y(t)
&=
-\dot X(t)
-2\gamma_0Y(t)
+2\gamma_0 k_{\rm B}T/\omega_0 .
\label{eq:varianze}
\end{align}

The initial thermal state at temperature \(T_\alpha\) imposes the conditions
\begin{equation}
X(0)=Y(0)=\frac{k_{\rm B}T_\alpha}{\omega_0},
\qquad
\dot X(0)=0 .
\end{equation}
Combining the two equations above, we obtain a third-order differential
equation for the position variance,
\begin{equation}
\dddot X(t)
+3\gamma_0\ddot X(t)
+(4\omega_0^2+2\gamma_0^2)\dot X(t)
+4\omega_0^2\gamma_0X(t)
=
4\gamma_0\omega_0 k_{\rm B}T .
\end{equation}
The general solution, including the stationary particular solution, is
\begin{equation}
X(t)
=
\frac{k_{\rm B}T}{\omega_0}
+
e^{-\gamma_0 t}
\left[
A+B\cos(2\eta t)+C\sin(2\eta t)
\right],
\end{equation}
where
\begin{equation}
\eta
=
\sqrt{\omega_0^2-\left(\frac{\gamma_0}{2}\right)^2}.
\end{equation}
Imposing the initial conditions, we finally obtain
\begin{equation}
X(t)=\langle Q^2(t)\rangle_{\rm ME}
=
\frac{k_{\rm B}T}{\omega_0}
+
e^{-\gamma_0 t}
\frac{k_{\rm B}(T_\alpha-T)}{\omega_0}
\frac{1}{\eta^2}
\left[
\omega_0^2
-\frac{\gamma_0^2}{4}\cos(2\eta t)
+\frac{\gamma_0\eta}{2}\sin(2\eta t)
\right].
\label{QME}
\end{equation}

Using the first of Eqs.~\eqref{eq:varianze}, the momentum variance is found to be
\begin{equation}
Y(t)=\langle P^2(t)\rangle_{\rm ME}
=
\frac{k_{\rm B}T}{\omega_0}
+
e^{-\gamma_0 t}
\frac{k_{\rm B}(T_\alpha-T)}{\omega_0}
\frac{1}{\eta^2}
\left[
\omega_0^2
-\frac{\gamma_0^2}{4}\cos(2\eta t)
-\frac{\gamma_0\eta}{2}\sin(2\eta t)
\right].
\label{PME}
\end{equation}

A direct comparison between Eqs.~\eqref{QME} and~\eqref{PME}, and
Eqs.~\eqref{sigmamarkov} and~\eqref{eq:Mmarkov}, respectively, using
\(\lambda=\gamma_0/2\), shows that the covariance matrix obtained from the
Born--Markov master equation coincides with that obtained from the
high-temperature, large-cutoff limit of the exact Caldeira--Leggett dynamics
in the absence of the initial-slip contribution.

Therefore, within the Born--Markov approximation, the relaxation dynamics
cannot display a Mpemba crossing. The same conclusion applies to more
restrictive Markovian descriptions, such as Lindblad master equations derived
after the rotating-wave approximation, since these approaches are based on
the same Markovian limit in which the initial-slip contribution is absent.

\section{Analytical evaluation of Matsubara sums} \label{sec:sums}
\label{thcontr}
In this Subsection, we provide the detailed evaluation of the following two relevant sums, that explain the results of $\langle \Delta Q(t)\rangle_{\mathrm{th}}$ and $\langle \Delta P(t)\rangle_{\mathrm{th}}$ presented in the main text.

\subsection{Evaluation of $S_1$}
We begin with the evaluation of 
\begin{equation}
S_1(x) = \sum_{n=0}^{\infty} \frac{L_n}{A_n - x}.
\label{sum}
\end{equation}
By inserting the explicit expression for $L_n$ from Eq.~\eqref{app:L0_definition}, we obtain:
\begin{equation}
S_1(x) = \frac{m \gamma_0 \omega_D^2}{2} \left[ \frac{1}{\omega_D - x} \cot\left(\frac{\beta \omega_D}{2}\right) + \frac{4}{\beta} \sum_{n=1}^{\infty} \frac{A_n}{(A_n^2 - \omega_D^2)(A_n - x)} \right].
\end{equation}
Using the definition of the Matsubara frequencies $A_n = 2\pi n / \beta$, the remaining infinite sum can be rewritten as:
\begin{equation}
\frac{4}{\beta} \sum_{n=1}^{\infty} \frac{A_n}{(A_n^2 - \omega_D^2)(A_n - x)} = \frac{\beta}{\pi^2} \sum_{n=1}^{\infty} \frac{n}{\left[ n - \frac{\beta x}{2\pi} \right] \left[ n^2 - \left(\frac{\beta\omega_D}{2\pi}\right)^2 \right]}.
\end{equation}
By performing a partial fraction decomposition and utilizing the identity for the digamma function $\psi(z)$,
\begin{equation}
\psi(y) - \psi(w) = \sum_{n=0}^{\infty} \left[ \frac{1}{n+w} - \frac{1}{n+y} \right],
\end{equation}
we can recast $S_1$ as
\begin{equation}
S_1(x) = \frac{m \gamma_0 \omega_D^2}{2\pi} \Bigg[ \frac{\pi}{\omega_D - x} \cot\left(\frac{\beta \omega_D}{2}\right)  
 + \frac{2x}{\omega_D^2 - x^2} \psi\left(-\frac{\beta x}{2\pi}\right) + \frac{1}{\omega_D + x} \psi\left(\frac{\beta\omega_D}{2\pi}\right) - \frac{1}{\omega_D - x} \psi\left(-\frac{\beta\omega_D}{2\pi}\right) \Bigg].
\end{equation}
Finally, using the reflection formula $\psi(-z) = \psi(z) + \frac{1}{z} + \pi \cot(\pi z)$, the sum $S_1(x)$ can be expressed in two equivalent forms:
\begin{equation}
\begin{split}
S_1(x) &= \frac{m \gamma_0 \omega_D}{\omega_D + x} \left\{ k_B T + \frac{x \omega_D}{\pi(\omega_D - x)} \left[ \pi \cot\!\left(\frac{\beta x}{2}\right) + \psi\left(\frac{\beta x}{2\pi}\right) - \psi\left(\frac{\beta\omega_D}{2\pi}\right) \right] \right\}\\
&= \frac{m \gamma_0 \omega_D}{\omega_D-x}
\left\{k_B\,T+\frac{\lambda_3\omega_D}{\pi(\omega_D+x)}\left[\psi\left(1-\frac{\beta x}{2\pi}\right)-
\psi\left(1+\frac{\beta\omega_D}{2\pi}\right)
\right]
\right\}.
\label{S1}
\end{split}
\end{equation}

\subsection{Evaluation of $S_2$}
We now evaluate the sum
\begin{equation}
S_2 = \sum_{j=1}^{3} \frac{\chi_j}{A_n - \lambda_j},
\label{sum2}
\end{equation}
and we show that it can be recast in the compact form
\begin{equation}
S_2 = \frac{\omega_D - A_n}{\mathcal{P}(A_n)},
\label{eq:S2}    
\end{equation}
where $\mathcal{P}(\lambda) = \lambda^3 + \omega_D\lambda^2 + (\omega_0^2 + \gamma_0\omega_D)\lambda + \omega_D\omega_0^2$ is the characteristic polynomial, whose roots $\lambda_j$ are assumed to have multiplicity one. 

We begin by writing $\mathcal{P}(\lambda) = \prod_{j=1}^{3}(\lambda - \lambda_j)$. Its derivative with respect to $\lambda$ is given by $\mathcal{P}'(\lambda) = \sum_{j=1}^{3} \prod_{j' \neq j}^{3}(\lambda - \lambda_{j'})$. Consequently, the weights $\chi_j$ are given by $\chi_j = (\omega_D - \lambda_j) \prod_{\ell \neq j} (\lambda_\ell - \lambda_j)^{-1}=({\omega_D - \lambda_j})/{\mathcal{P}'(\lambda_j)}$, allowing us to express $S_2$ as
\begin{equation}
S_2 = \sum_{j=1}^{3} \frac{\omega_D - \lambda_j}{\mathcal{P}'(\lambda_j)} \frac{1}{A_n - \lambda_j}.
\label{eq:S2rewr}  
\end{equation}
Consider now the rational function
$
({\omega_D - A_n})/{\mathcal{P}(A_n)}.
$
Its partial fraction decomposition yields
\begin{equation}
\frac{\omega_D - A_n}{\mathcal{P}(A_n)} = \sum_{j=1}^{3} \frac{c_j}{A_n - \lambda_j}, 
\label{eq:partialdecomp}
\end{equation}
{with}  $c_j = ({\omega_D - \lambda_j})/{\mathcal{P}'(\lambda_j)}.$
Comparing Eq.~\eqref{eq:S2rewr} with Eq.~\eqref{eq:partialdecomp}, it is immediate that Eq.~\eqref{eq:S2} holds.

\FloatBarrier
\bibliography{references}